\documentclass[pre,a4paper,12pt, preprintnumbers,floatfix,english]{revtex4}

\usepackage[utf8]{inputenc}

\usepackage[T1]{fontenc}
\usepackage{babel}
\usepackage{fancyhdr}
\usepackage{ifpdf}
\usepackage{graphicx}
\usepackage{color}
\usepackage{lineno}
\usepackage{pslatex}
\usepackage[justification=Justified,
   format=plain]{caption}
\usepackage{subcaption}
\usepackage{amsmath}

\usepackage{bm}
\usepackage{dcolumn}

\usepackage{amsmath}
\usepackage{amssymb}
\usepackage{amscd}

\ifpdf
\usepackage{epstopdf}
\usepackage[%
  pdfauthor={},%
  pdfstartview=FitH,%
  bookmarks=true,%
  bookmarksopen=true,%
  breaklinks=true,%
  colorlinks=true,%
  linkcolor=blue,anchorcolor=blue,%
  citecolor=blue,filecolor=blue,%
  menucolor=blue,pagecolor=blue,%
  urlcolor=blue]{hyperref}
\else
\usepackage[%
  pdftitle={Notes on As-S melts, Matthieu},%
  pdfauthor={Notes on As-S melts, Matthieu},%
  pdfsubject={lGe2Se3},%
  pdfkeywords={lGe3Se3},%
  breaklinks=true,%
  colorlinks=true,%
  linkcolor=blue,anchorcolor=blue,%
  citecolor=blue,filecolor=blue,%
  menucolor=blue,pagecolor=blue,%
  urlcolor=blue]{hyperref}
\fi

\begin{document}
\include{bibsymb}

\draft

\title{Intrinsic Limitation of Conductivity in Depolymerized Sodium-ion Glassy Networks}

\author{L. Legrand, L.-M. Poitras, N. Sator, M. Micoulaut\footnote{email: matthieu.micoulaut@sorbonne-universite.fr}}

\affiliation{Sorbonne Université, Laboratoire de Physique Théorique de la Matière Condensée, CNRS UMR 7600, 4 Place Jussieu, 75252 Paris Cedex 05, France}

\date{\today}

\begin{abstract} 
\textbf{Abstract:}  
The electric properties of a model fast-ion electrolyte glass (100-$x$)SiS$_2$ -- $x$Na$_2$S are investigated by means of classical molecular dynamics simulations. These materials are promising candidates for battery applications and the conductivity is thought to be essentially driven by the concentration of Na charge carriers. We first set up a Buckingham-Coulomb type potential able to describe the atomic structure and experimental structure functions (structure factor) in an improved fashion with respect to previous reported force fields. A systematic investigation of properties with modifier content Na$_2$S (50~\%$\leq x\leq$80~\%) leads to an unexpected result, that is, a near constant behavior of the conductivity $\sigma(x)$ with Na$_2$S increase for various isotherms in the liquid state. The analysis indicates that unlike Li-based electrolytes, the diffusivity difference between network (Si,S) and modifier (Na) species is small, and this leads to a contribution to conductivity dominated by (Si,S) atoms. While the concentration of network species in the range 66~\%$\leq x\leq$80~\% decreases, no dramatic increase in Na diffusivity is obtained, and the emergence of molecular Na$_2$S in the structure at large modifier content also induces profound structural changes. Unlike Lithium glassy electrolytes, the design of Na-based batteries must, therefore, take into account the contribution of the network species.

\end{abstract}

\maketitle

\section{Introduction}

At a time when modern human activity has an ever-increasing demand for energy, and the technological progress that has enabled it to flourish has also made it dependent on highly polluting fossil fuels, the disruption of biodiversity and climate should prompt us to reconsider the direction of technological innovation. Electrical energy generated by non-carbon energies such as solar, wind, hydro and even nuclear power are good alternatives to fossil fuel combustion. As a result, electric energy and the corresponding energy storage technologies using next-generation batteries will play an essential role in the advent of this sustainable future. Despite the current success of the Li-ion battery technology, the carbon footprint during manufacture, the instability of its flammable liquid electrolyte, and the dependence on lithium resources are leading us to look for greener and safer technologies\cite{tarascon_nature_2001, weiss_aem_2021}. All-solid-state batteries using sodium represent, therefore, a promising alternative. However, to make it attractive, a number of conditions need to be met, such as good ionic conductivity within the electrolyte as well as high mechanical and chemical stability of the electrode/electrolyte interface\cite{huang_ce_2024}.
\par
In this respect, amorphous or glassy electrolytes \cite{grady_frontiers_2020} are seen as promising candidates to reach efficiency criteria for battery applications due to the possibility of combining a nearly infinite number of components in the base materials, enabling the continuous improvement of crucial properties such as ionic conduction or mechanical properties. More precisely, the increased polarization of sulfur with respect to oxygen makes sulfide glasses \cite{pradel_cras_2022} attractive, and these display a significant level of conductivity in highly depolymerized materials, e.g. 10$^{-3}$~$\Omega$$^{-1}$cm$^{-1}$ in 3Li$_2$S -- P$_2$S$_5$ glasses\cite{eckert_cm_1990,mizuno_am_2005,mori_cpl_2013,ohara_sr_2016}. Among such sulfides, the Na$_2$S -- SiS$_2$ system containing a network former (SiS$_2$) and a modifier (Na$_2$S) displays a more reduced level of conductivity\cite{pradel_cras_2022} (2$\times$10$^{-6}~\Omega^{-1}$ cm$^{-1}$) compared to its lithium counterpart\cite{pradel_jncs_1995} (10$^{-4}$~$\Omega$$^{-1}$ cm$^{-1}$). In the present system, the effect of the alkali sulfide is similar to the one encountered in oxides and in all modified glasses using a network former (GeO$_2$, P$_2$S$_5$, B$_2$O$_3$,...), i.e. the modifier (Na$_2$S) will disrupt the base network made of SiS$_2$ tetrahedra linked by corners and edges\cite{micoulaut_ropp_2016}, and will lead to a global depolymerization of the base network structure by the conversion of bridging sulfur (BS, i.e. sulfur atoms connecting two tetrahedra) into so-called "non-bridging" sulfur (NBS) close to which alkali ions are present. A convenient way to quantify this conversion builds on the so-called $Q^n$ speciation which enumerates the number $n$ of bridging sulfur (BS) atoms among a given SiS$_{4/2}$ tetrahedron. At elevated modifier composition which leads to the most elevated conductivities, the structure is thought to be made of a majority of isolated tetrahedra Na$_4$SiS$_4$ corresponding to possible 100~\% so-called $Q^0$ species where the superscript indicates the number of BS atoms.
\par
In order to improve the conductivity level of such Na-based sulfide glasses\cite{micoulaut_jncs_2024, micoulaut_prb_2023}, a straightforward way is to modify the composition $x$ of $x$Na$_2$S -- (1-$x$)SiS$_2$ glasses in order to increase the amount of charge carriers (Na). This strategy is often applied albeit limited by the glass-forming domain (GFD) of the considered system. New synthesis techniques (e.g. ball-milling) can eventually extend the GFD, and produce glass compositions with significantly higher amounts of certain elements. For instance, it is now possible to obtain glasses with up to 80~\% modifier such as 80Na$_2$S--20Ga$_2$S$_3$\cite{denoue_mrb_2021}. It is, therefore, tempting to investigate the properties of such highly depolymerized glasses. 
\par
The present study focuses on ultra-depolymerized networks using atomic scale simulations in order to link aspects of structure with dynamic and electrical properties by focusing on the effect of composition on xNa$_{2}$S-(1-x)SiS$_{2}$ studied from $x$=50\% to x=80\%. The choice is motivated by the fact that the physical properties of such glasses at such high modifier content have been rarely studied. Unexpectedly, we find that for $x\geq$50~\% the increase of conductivity with composition $x$ is weak. The reason comes from a too small diffusivity difference $\Delta D$ between Na and the network species, which all contribute to the conductivity. This situation is at variance with Li-based glasses where such difference $\Delta D$ is at least one order of magnitude larger, and makes the conductivity contribution of Li dominant. The result has broader consequences as it suggests that, unlike for Li-based all solid state batteries (ASSB), the design of future Na-based ASSB must also consider the crucial role of the network species. 
\section{Simulation details}
\subsection{General framework}

Classical molecular dynamics (MD) simulations were conducted in NVT ensemble on $N$=3000 particle glassy systems containing various concentrations of network modifier Na$_2$S. The starting configuration was obtained from a previous converged configuration of 50Li$_2$S--50SiS$_2$ \cite{poitras_prb_2023} where the Li atoms were substituted by Na atoms, and the size of the simulation box adjusted in order to meet the experimental\cite{watson_jpcb_2018} glass density of 0.0430~\AA$^{-3}$. The equations of motion were integrated using the Verlet algorithm with a time step of 1~fs. The system was maintained at 2000~K for 100~ps, followed by a quench at a cooling rate of 1~K/ps to 300~K in NVT ensemble using a Nosé-Hoover thermostat. Finally, statistical analyses of the glasses were carried out in the NVT ensemble over 10~ns.
\par
To model xNa$_{2}$S-(1-x)SiS$_{2}$ glasses, we used the Buckingham potential which includes a short-range repulsive interaction, a Coulomb interaction, and a long-range attractive dispersive interaction:

\color{black}
\begin{equation}
\label{buck}
    V_{ij}(r)=A_{ij}\exp\biggl({-r/\rho_{ij}}\biggr)+\frac{q_iq_j}{4\pi \epsilon_0 r}-\frac{C_{ij}}{r^6},
\end{equation}
where $r$ is interatomic distance, $A_{ij}$, $\rho_{ij}$, and $C_{ij}$ are parameters, $q_i$ are the charges (in $e$ units), and $\epsilon_0$ is the permittivity of vacuum. $i$ and $j$ refer to the type of two distinct atoms. 
\par
In order to fix the parameters of eq. (\ref{buck}), using various amorphous structures obtained after a melt-quench procedure from molecular dynamics simulations, we minimize a goodness-of-fit (Wright) parameter\cite{wright_jncs_1993} that builds on a direct comparison between the experimental\cite{dive_jpcb_2018} and the simulated data. 
\begin{eqnarray}
\label{acw}
    R_X=\Bigg(
    {\frac {\sum\limits_i \Big[S_{exp}(k_i)-S_{calc}(k_i)\Big]^2}
    {\sum\limits_i S^2_{exp}(k_i)}}
    \Bigg)^{1/2}
\end{eqnarray}
The fit stops as $R_X$ becomes lower than a certain value that we have fixed to 0.005, which is a typical value for concluding reproductions of structure factors\cite{flores_prb_2018}. In order to reproduce correctly the region at low momentum transfer, it is necessary to incorporate a three-body interaction that applies specifically to the triplet Si-S-Si involving a bridging sulfur (BS) atom

\begin{eqnarray} \label{3body} 
V_3(\theta)=k_b(\theta-\theta_0)^2,
\end{eqnarray}

and only active beyond a certain cut-off value (2.7~\AA) corresponding to the minimum of the Si--S pair correlation function\cite{dive_jpcb_2018}. The inclusion of such a 3-body term that constrains the angular excursion around the BS atom within the network has been shown to be important\cite{bertani_jacers_2022,bertani_prm_2021} in corresponding oxides in order to reproduce a certain number of features among which, the first sharp diffraction peak observed at low momentum transfer $k$ in the structure factor $S(k)$. In sulfide glasses, the addition of the interaction in Eq. (\ref{3body}) becomes even more critical because sulfide networks exhibit both corner-sharing (CS) and edge-sharing (ES) polyhedral connections having a distinct average angle around a BS atom. These mixed connections create a bimodal bond angle distribution such as Ge-S-Ge in e.g. GeS$_2$ network formers\cite{chakraborty_prb_2017,micoulaut_prb_2022}, and cannot be reproduced from an interaction restricted to the single two-body interaction of Eq. (\ref{buck}).

\section{Results}

\subsection{Validation of the force-field}
\begin{figure}
\includegraphics[width=0.5\linewidth]{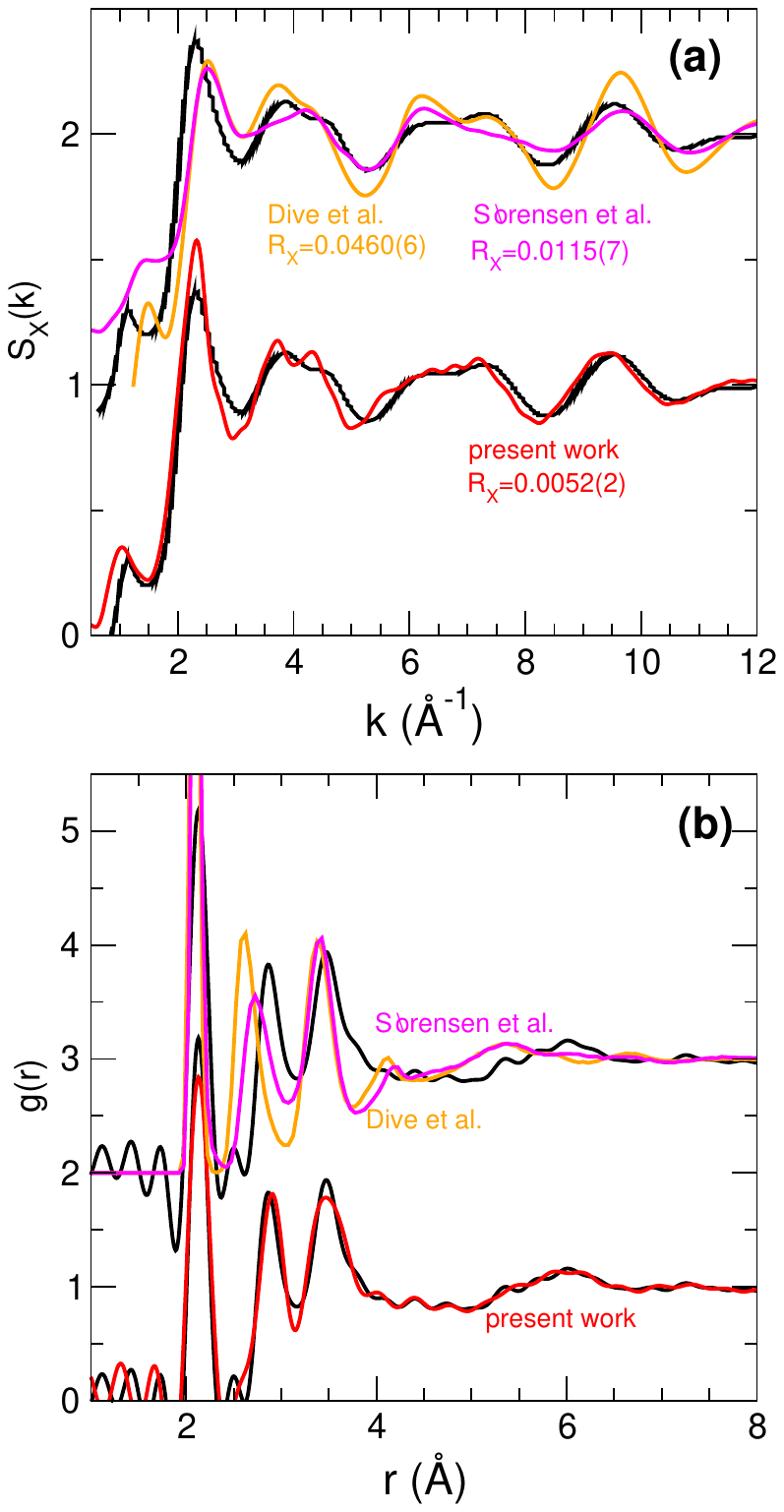}
\caption{\label{sq} Calculated (red) X-ray weighted structure factor $S_X(k)$ (a) and pair correlation function $g(r)$ (b) of 50Na$_2$S -- 50SiS$_2$, compared to experimental data from Dive et al. \cite{dive_jpcb_2018} (black, duplicated) and to results from previous force-fields from Dive et al. \cite{dive_jpcb_2018} (orange) and S\o rensen et al. \cite{sorensen_jpcb_2023} (magenta). Note that the total experimental $g(r)$ (not reported in Ref. \onlinecite{dive_jpcb_2018}) been obtained from a Fourier transform of the experimental $S_X(k)$. The Wright parameter $R_X$ is indicated and has been calculated in the range 0.6~\AA$^{-1}\leq k\leq$12.0~\AA$^{-1}$.
 }
\end{figure}

We first verify that the parametrized force-field is able to reproduce measured structure functions from X-ray scattering measurements. Fig. \ref{sq} represents the calculated (red) X-ray weighted structure factor $S_X(k)$ (panel a) and pair correlation function $g(r)$ (panel b) that are compared to experimental data\cite{dive_jpcb_2018} (black).
\par
One obviously acknowledges a substantial improvement with respect to results from previous molecular molecular dynamics simulations\cite{dive_jpcb_2018,sorensen_jpcb_2023} (magenta and orange curves) in both the real and reciprocal space. It is important to stress that the present parametrized  force-field now captures all important features of the experimental structure factor $S_X(k)$, i.e. the principal peaks measured at 2.5~\AA$^{-1}$, and 4~\AA$^{-1}$ but also the first sharp diffraction peak measured\cite{dive_jpcb_2018} at 1.13~\AA$^{-1}$. Our results are at variance with the previous MD simulations which contained obvious flaws such as i) an obvious shift to large $k$ in the first principal peak position, ii) a poor reproduction of all the other peaks and iii) a phase-lag at large momentum transfer indicative of possible a spurious short-range order in real space. The calculation of the corresponding Wright parameter (eq. (\ref{acw})) provides some quantitative information on the level of agreement since we find $R_X$=0.0052 when we compare $S_{calc}$ and $S_{exp}$ in the range range 0.6~\AA$^{-1}\leq k\leq$12.0~\AA$^{-1}$ for the present force-field. The two previous force fields lead to $R_X$=0.0115\cite{sorensen_jpcb_2023} and 0.0460\cite{dive_jpcb_2018}, respectively.
\par
Once Fourier transformed, the spurious effects observed at large $k$ in the previous simulations\cite{dive_jpcb_2018,sorensen_jpcb_2023} manifest in Fig. \ref{sq}b by an incorrect second- and third-shell correlation of neighbors supposed to be found experimentally\cite{dive_jpcb_2018} at 2.89~\AA\ and 3.50~\AA. Conversely, the present force field (eq.(\ref{buck})) is able to reproduce all features of the pair correlation function $g(r)$, and these essentially contain a first principal peak at 2.13~\AA\ (experimentally\cite{dive_jpcb_2018} 2.14~\AA) arising from the tetrahedra apex Si-S, a secondary peak at 2.91~\AA\ (experimentally 2.88~\AA) related mostly to Na-S correlations, and finally a third peak at 3.47~\AA\ (experimentally 3.47~\AA) which is due to multiple contributions (the S-S tetrahedral edge, Na-Na and Si-Na). All simulated peak characteristics are now obviously reproduced, i.e. amplitudes, positions and widths with minor discrepancies so that the present force-field represents a clear and unambiguous improvement with respect to previous simulation efforts. In addition (see below), the resulting structure is found to contain both corner-sharing (CS) and edge-sharing (ES) tetrahedral motifs, and this structural feature represents another important improvement with respect to the previous MD studies\cite{dive_jpcb_2018,sorensen_jpcb_2023}.

\subsection{Diffusivity and conductivity of NS melts}

\begin{figure}
\includegraphics[width=0.7\linewidth]{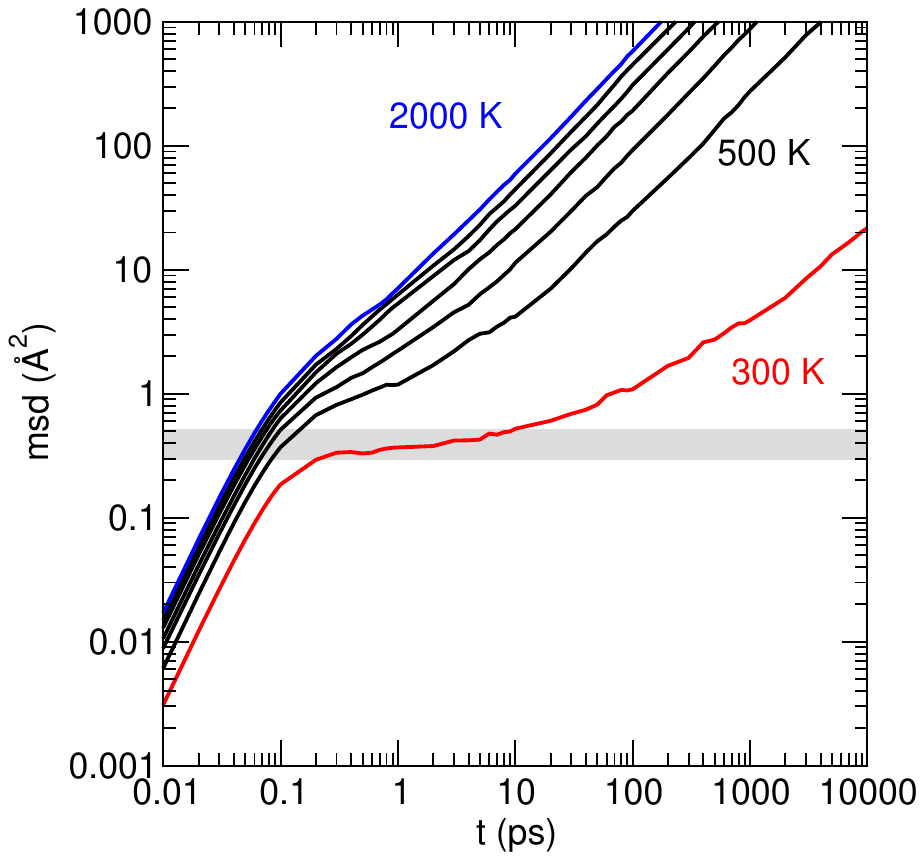}
\caption{\label{msd} Na mean square displacement $\langle r^2_{Na}(t)\rangle$ for different isotherms in NS melts : 2000~K (blue), 1750~K, 1500~K, 1250~K, 1000~K, 500~K and 300~K (red).
}
\end{figure}

In order to calculate dynamic and electric properties, we focus on the mean square displacement (msd, Fig. \ref{msd}) of the atoms in melts at different temperatures, which is defined from the positions ${\bf r}_j(t)$ at time $t$ 
\begin{equation}\label{msdtoD}
   \langle r_k^2(t)\rangle = \biggl\langle \frac{1}{N_k} \sum_{j=1}^{N_k} | \textbf{r}_j(t)-\textbf{r}_j(0)|^2 \biggr\rangle
\end{equation}
\noindent
where the sum is taken over all atoms of type $k$ ($k$=Si,S,Na). The self-diffusion $D_k$ constant in the long-time limit is then defined from the Einstein equation:
\begin{eqnarray}\label{eq.msddiff}
    D_k=\frac{1}{6} \lim_{t\to\infty} \frac{\text{d}\langle r_k^2(t)\rangle}{\text{d}t}.
\end{eqnarray}

\begin{figure}
\begin{center}
\includegraphics[width=0.7\linewidth]{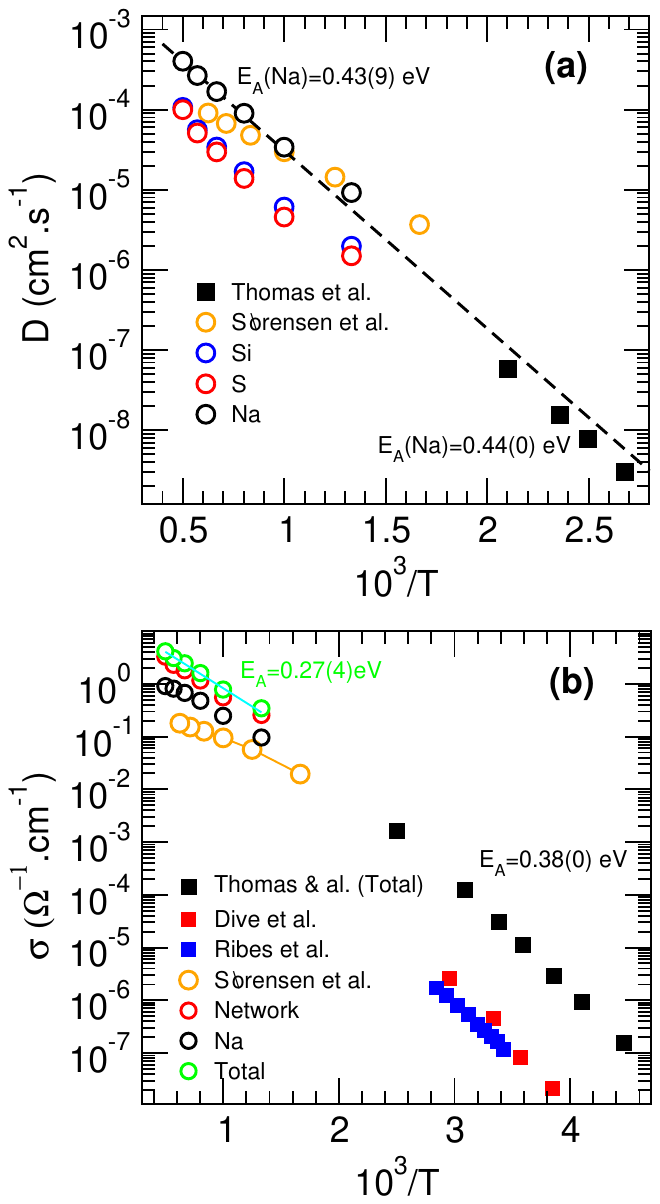}    
\end{center}
\caption{\label{diffus} a) Calculated diffusion coefficient $D$ in NS melts as a function of inverse temperature, compared to data from $^{22}$Na tracer experiments in 56Na$_2$S -- 44SiS$_2$ of Thomas et al. \cite{thomas_jacer_1985}, and to MD simulations using a different force-field \cite{sorensen_jpcb_2023}. b) Calculated conductivity (green symbols) in NS melts as a function of inverse temperature, compared to experimental data measured in the glassy state from Ribes et al. \cite{ribes_jncs_1980}, Dive et al. \cite{dive_jpcb_2018}, and Thomas et al. \cite{thomas_jacer_1985}, and to MD data from S\o rensen et al.\cite{sorensen_jpcb_2023}. Note that the set of data of Thomas et al. corresponds to a 56Na$_2$S -- 44SiS$_2$ composition, as in panel a. The contribution of the network (Si,S, open red circles) to the conductivity is separated from the one due to sodium (black open circles).
}
\end{figure}

Diffusivity results of the NS melt now appear in Fig. \ref{diffus}, and follow an Arrhenius behavior $D\propto\exp[-E_A/k_BT]$ for all species with activation energies $E_A$ found to be 0.43(9), 0.64(7) and 0.60(0)~eV for Na, S and Si, respectively. Here the semi-log representation as a function of inverse temperature obviously reveals a dynamics of activated (Arrhenius) type. These values $E_A$ are somewhat higher than those previously determined (e.g. 0.25(8)~eV for Na\cite{sorensen_jpcb_2023} to be compared with our 0.43(9)~eV) but the latter might be altered by the wrong structural properties as discussed above. The present calculated activation energy for Na diffusion is highly consistent with the one determined by Thomas et al. at lower temperatures from $^{22}$Na tracer diffusion experiments\cite{thomas_jacer_1985} on a slightly different composition (56Na$_2$S -- 44SiS$_2$). Here, the measured Na activation energy E$_A$ was found to be 0.44(0)~eV, i.e. identical to our calculated $E_A$=0.43(9). The force-field (eq. (\ref{buck})), thus, appears to be extremely reliable as it describes correctly both the structure (Fig. \ref{sq}) and the dynamics (Fig. \ref{diffus}). For obvious numerical limitations, we are not able to explore numerically the diffusivity in the same temperature range as Thomas et al.\cite{thomas_jacer_1985} because of the onset of glassy behavior that limits the motion of the particles at low temperature within the computer timescale, and does not permit to obtain the diffusive regime at ambient conditions. This is clearly visible from the 300~K msd profile (Fig. \ref{msd}, red curve) which displays the usual cage-like motion (here at msd$\simeq$0.5~\AA$^2$ between 0.5-20~ps) prior to a reduced dynamics at long times (10~ns).
\par 
In the present sodium system, the diffusivity difference $\Delta D$ between the modifier (Na) and the network species (Si,S) is smaller than in the corresponding lithium counterpart\cite{micoulaut_jncs_2024} as we calculated at e.g. $T$=1000~K, $D_{Na}$=3.54$\times$10$^{-5}$~cm$^2$.s$^{-1}$ and for the network (N) species $D_{N}\simeq$4.49-5.62$\times$10$^{-6}$~cm$^2$.s$^{-1}$, i.e. a ratio $D_{Na}/D_{N}\simeq$7. This ratio is substantially smaller as we found $D_{Li}/D_{N}\simeq$100 in the Li-based electrolyte at the same temperature\cite{micoulaut_jncs_2024}, and indicative of an obvious decoupling between the alkali and the network dynamics in Li thiosilicates. In the sodium glass, this "small" difference $\Delta D$ between $D_{Na}$ and $D_{N}$ will have profound consequences on the conductivity behavior, as discussed below. 

The conductivity ($\sigma$) as a function of temperature $T$ can be obtained from the Green-Kubo equation \cite{hansen_book}:
\begin{equation}\label{nernsteinsteineq}
    \sigma(T)=\lim_{t\to\infty} {\frac {Ne^2}{6tVk_BT}}\sum_{\lim\limits_{i,j}}z_izj
    \biggl\langle [{\bf r}_i(t)-{\bf r}_i(0)][{\bf r}_j(t)-{\bf r}_j(0)]\biggr\rangle
\end{equation}
where $V$ is the volume of the simulation box, $e$ is the elementary charge, $z_i$ and $z_j$ are the fractional charges of ions $i$ and $j$ of the interaction potential (Table \ref{tab_nasis}), respectively. Here ${\bf r}_i(t)$ is the position of atom $i$, and the brackets $\langle\rangle$ denote ensemble averages. The calculated conductivity $\sigma(T)$ (eq. (\ref{nernsteinsteineq})) for the NS melts is displayed in Fig. \ref{diffus}b (green symbols) and compared to different experimental measurements in the glass\cite{ribes_jncs_1980,dive_jpcb_2018,thomas_jacer_1985} and to a previously\cite{sorensen_jpcb_2023} calculated $\sigma(T)$ from a slightly different force-field. As for the diffusivity, we note that the calculated $\sigma(T)$ behavior in the liquid state is compatible with these experimental measurements performed at 10$^3/T>$2.5 since the trend at high temperature eventually extrapolates to the data at low temperature. Both sets display an obvious Arrhenius with activation energies found to be 0.38(0)~eV and 0.27(4)~eV for experiment and simulation, respectively. These differences might be the signature of a so-called Arrhenius crossover in the glass transition region (here $T_g$=545~K for the NS glass\cite{pradel_cras_2022}), which reflects the fact that, similarly to other glass-forming electrolytes\cite{golodnitsky_jes_2015,tu_m_2022, aniya_pssb_2020,malki_gca_2015,fan_jncs_2020}, conductivity is enhanced once the underlying network softens with the reduction of viscosity, and might eventually lead to conductivity jumps.
\par
In order to have more details on the role of the different species, we split the total $\sigma(T)$ into contributions from the Na ions, and from the network (N) species (Si,S), i.e. one focuses on different contributions $\sigma_k$ ($k$=Na, N) :
\begin{eqnarray}
    \sigma_k={\frac {e^2}{Vk_BT}z_k^2D_k(T)}
\end{eqnarray}
where $D_k$ is the self-diffusion constant of the species $k$ determined by eq. (\ref{eq.msddiff}), and $z_k$ are the fractional charges (Table \ref{tab_nasis}). Here we note (Fig. \ref{diffus}b) that the overall conductivity is dominated by the network species (Si,S, red circles) up to high temperature, and at the lowest investigated temperature (750~K) the ratio between network and ion conttribution $\sigma_N/\sigma_{Na}$ is of about 2.7. This situation is at variance with the one encountered in the lithium counterpart\cite{micoulaut_jncs_2024} where the important diffusivity difference between Li and N species leads e.g. to $\sigma_N/\sigma_{Li}\simeq$0.12 at 900~K, i.e. $\sigma_{Li}$ dominates the conductivity of the lithium thiosilicates. These features highlight the role of structure, and also the role of the size of the alkali modifier on the overall behavior of electric properties.

\subsection{Diffusivity and conductivity of highly depolymzerized networks}

Having validated both the structure and the dynamics of the Buckingham force-field (eq. (\ref{buck})) with respect to experiments for the single composition $x$=50~\%, we now investigate the electric properties in the very limit of the glass-forming domain which is known\cite{pradel_cras_2022,watson_jncs_2017} to be of about 66-70~\%. Thanks to a new technique building on a mechanical synthesis route (ball milling), the GFD can now be extended to about 75~\% modifier as in recent studies in e.g. lithium thiophosphates\cite{dietrich_jmca_2017}. Importantly, it should be stressed that the stoichiometry of the network former fixes a theoretical limit to network depolymerization that is entirely achieved once the base polytopes (e.g. the SiS$_4$ tetrahedra in the present case) are entirely disconnected from the rest of the network. In the present alkali thiosilicates, this limit is reached at 66~\% modifier which corresponds to a network made of isolated (M=Li,Na) M$_4$SiS$_4$ tetrahedra (so-called $Q^0$ units\cite{pradel_jncs_1995}) containing no longer BS atoms.
\par

\begin{figure}
\includegraphics[width=0.7\linewidth]{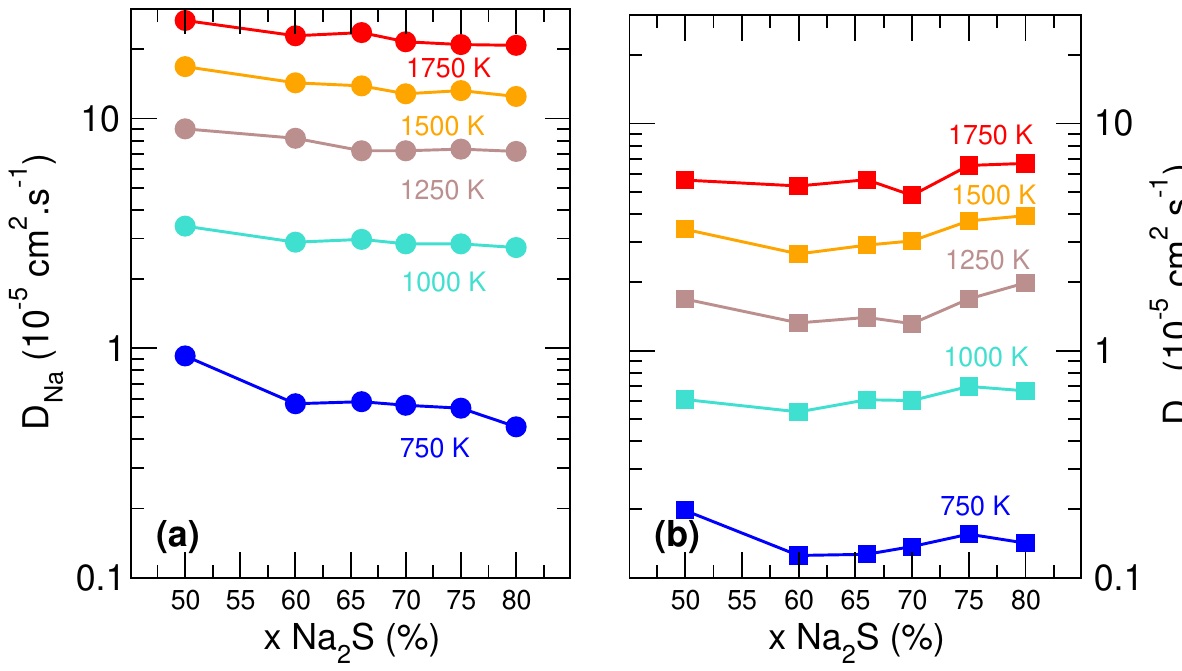}
\caption{\label{diff1} Calculated diffusivity $D_k(T)$ in $x$ Na$_2$S -- (1-$x$)SiS$_2$ melts for different isotherms for Na (a) and network species (S,Si) (b). Error bars are of the size of the symbols.}
\end{figure}

\begin{figure}
\includegraphics[width=0.7\linewidth]{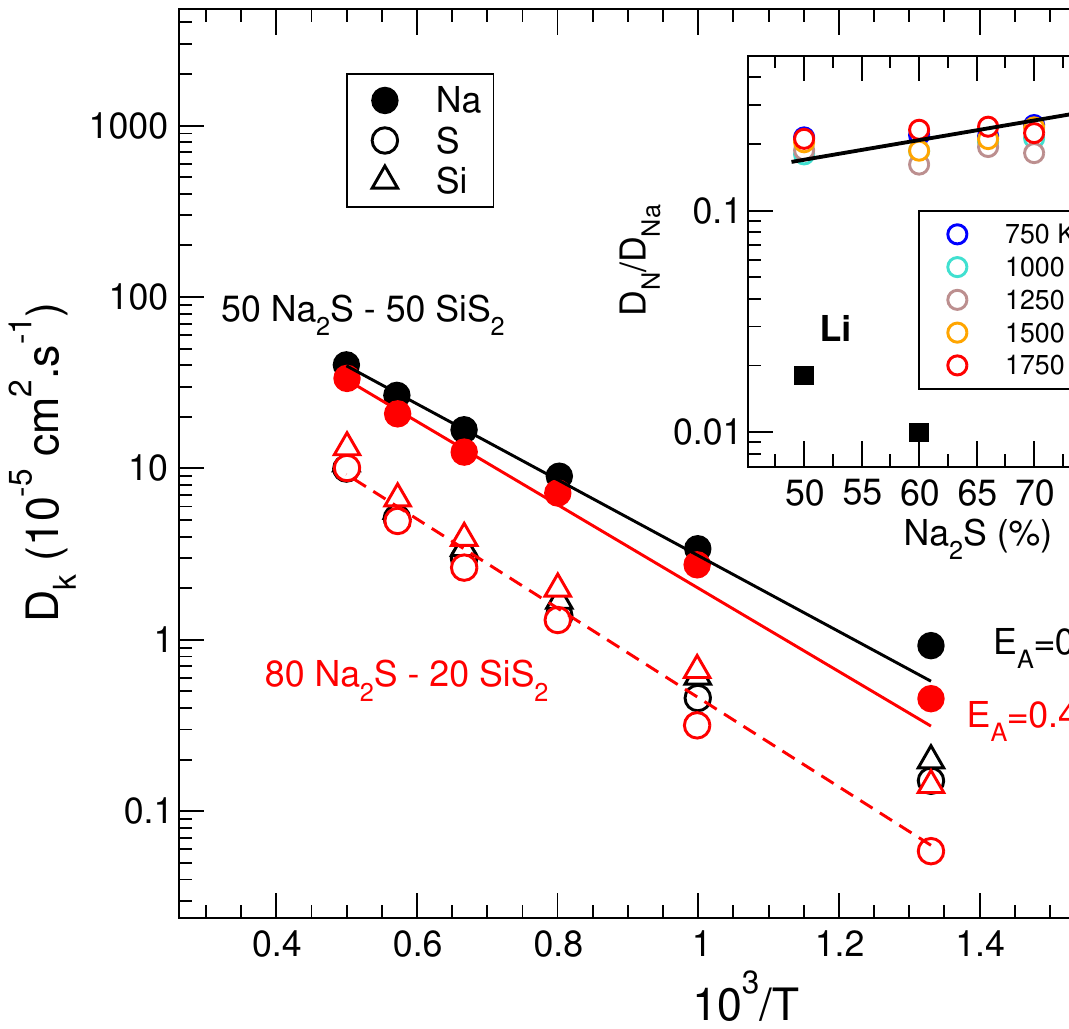}
\caption{\label{diff2} Calculated diffusivity $D_k(T)$ in 50Na$_2$S -- 50SiS$_2$ (black) and 80Na$_2$S -- 20SiS$_2$ (red) with Arrhenius fits (solid and broken lines). The inset shows the Na versus network species diffusivity ratio $D_{N}/D_{Na}$ as a function of modifier concentration $x$ Na$_2$S for the different considered isotherms. For comparison, the ratio $D_{N}/D_{Li}$ is represented (filled squares) for 50Li$_2$S -- 50SiS$_2$\cite{micoulaut_jacers_2024} and 60Li$_2$S -- 40SiS$_2$\cite{micoulaut_jncs_2024}.}
\end{figure}

Figure \ref{diff1} displays the behavior of Na and network species (Si,S) diffusivities for different isotherms as a function of modifier content. As for the 50:50 compound, the dynamics between the Na ions and the network species is somehow decoupled but, again, with a difference that is substantially smaller than for the Li counterpart\cite{micoulaut_jacers_2024,micoulaut_jncs_2024}. Noteworthy is the fact that with increasing Na content, one does not acknowledge an increase of diffusivity and this underscores the fact that the effect of depolymerization on the dynamics is weak, i.e. between the 50:50 and the 80:20 compound, there is only a slight increase of $D_k$ for the network species and a near constant value for $D_{Na}$ which is found to be of the order of 3.0$\times$10$^{-5}$~cm$^2$.s$^{-1}$ at 1000~K (Fig. \ref{diff1}a). We rule out possible effects of glass transition temperature ($T_g$) dependence with modifier content $x$ since it has been reported that this dependence is small, i.e. of about\cite{pradel_cras_2022} 270$^o$C between $x$=50~\% and 60~\%, and 247$^o$C for x=66~\%\cite{watson_jpcb_2018}. In addition, the temperature evolution between the two end compositions (50 and 80~\%) appears to be rather small (Fig. \ref{diff2}) and this is highlighted in an Arrhenius representation where a small decrease in Na diffusivity is acknowledged at the lowest temperature for the 80:20 composition. A corresponding Arrhenius fit using $D\propto\exp[-E_A/k_BT]$ shows that the activation energy for diffusion $E_A$ increases from 0.43(8)~eV to 0.48(2)~eV from $x$=50~\% to 80~\%, also indicative for a reduced dynamics at elevated modifier content. 
\par
The respective evolution of network dynamics versus alkali dynamics is detected in the inset of Fig. \ref{diff2} which represents the ratio between network and Na diffusivity $D_N/D_{Na}$. For all temperatures, this ratio is found to increase, i.e. the diffusion of the network species increases faster than $D_{Na}$ with modifier content, as already reflected in Fig. \ref{diff1}b, and this underscores the fact that the depolymerization induces an increased dynamics for the network species. The situation is, again, at variance with the Li counterpart for which the corresponding ratio $D_N/D_{Li}$ is found to decrease\cite{micoulaut_jacers_2024,micoulaut_jncs_2024} with $x$, i.e. the contribution of the network which is already small in the 50:50 lithium thiosilicates becomes even smaller at large modifier content. 

\begin{figure}
\includegraphics[width=0.7\linewidth]{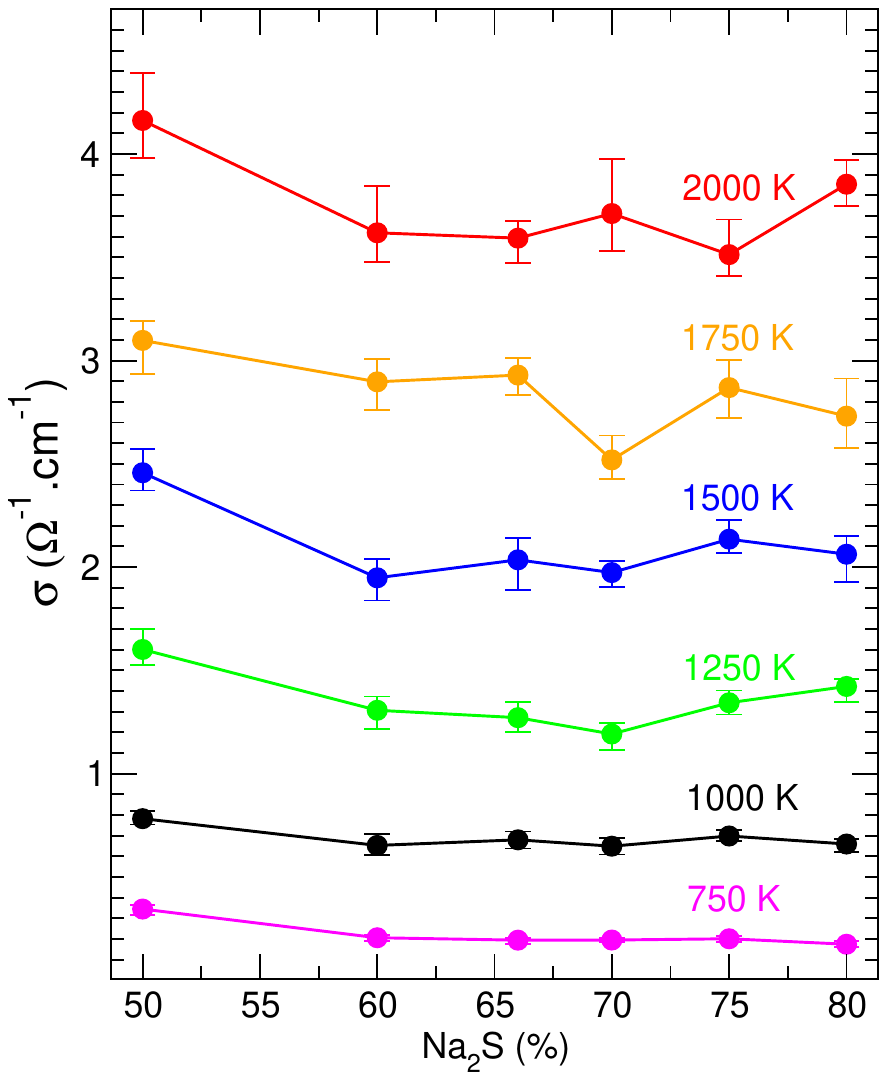}
\caption{\label{condux_vs_x} Calculated system conductivity $\sigma(T)$ as a function of modifier concentration $x$ Na$_2$S in $x$Na$_2$S -- (1-$x$)SiS$_2$ melts for different isotherms.
}
\end{figure}

Similarly, Figure \ref{condux_vs_x} now represents the calculated conductivity of melts with different temperatures as a function of modifier concentration $x$. Importantly, we remark again that $\sigma(T)$ remains almost constant over the entire concentration domain or that, at least, there is no significant evolution with $Na_2$S.

\section{Discussion}
The evolution of the electric properties is linked with a profound alteration of the network structure. For the NS glass, it is well known that the short-range connectivity is realized by corner- (CS) and edge-sharing (ES) SiS$_{4/2}$ tetrahedral connections. These are usually revealed from $^{29}$Si Nuclear Magnetic Resonance (NMR), and results\cite{eckert_jncs_1989,watson_jncs_2017,pradel_cras_2022} for a series of alkali modified thiosilicates glasses have shown that the fraction of ES units decreases continuously with modifier content. For a 50:50 ratio (i.e. Na$_2$SiS$_3$ or Li$_2$SiS$_3$), the proportion of edge-sharing tetrahedra is about 23~\% \cite{eckert_jncs_1989,watson_jncs_2017} and this indicates that the glass structure maintains a partial chain-like structure with the base tetrahedra SiS$_{4/2}$ being connected both by edges and corners. 

\subsection{Breakdown of ES connectivity}

\begin{figure}
\includegraphics[width=0.7\linewidth]{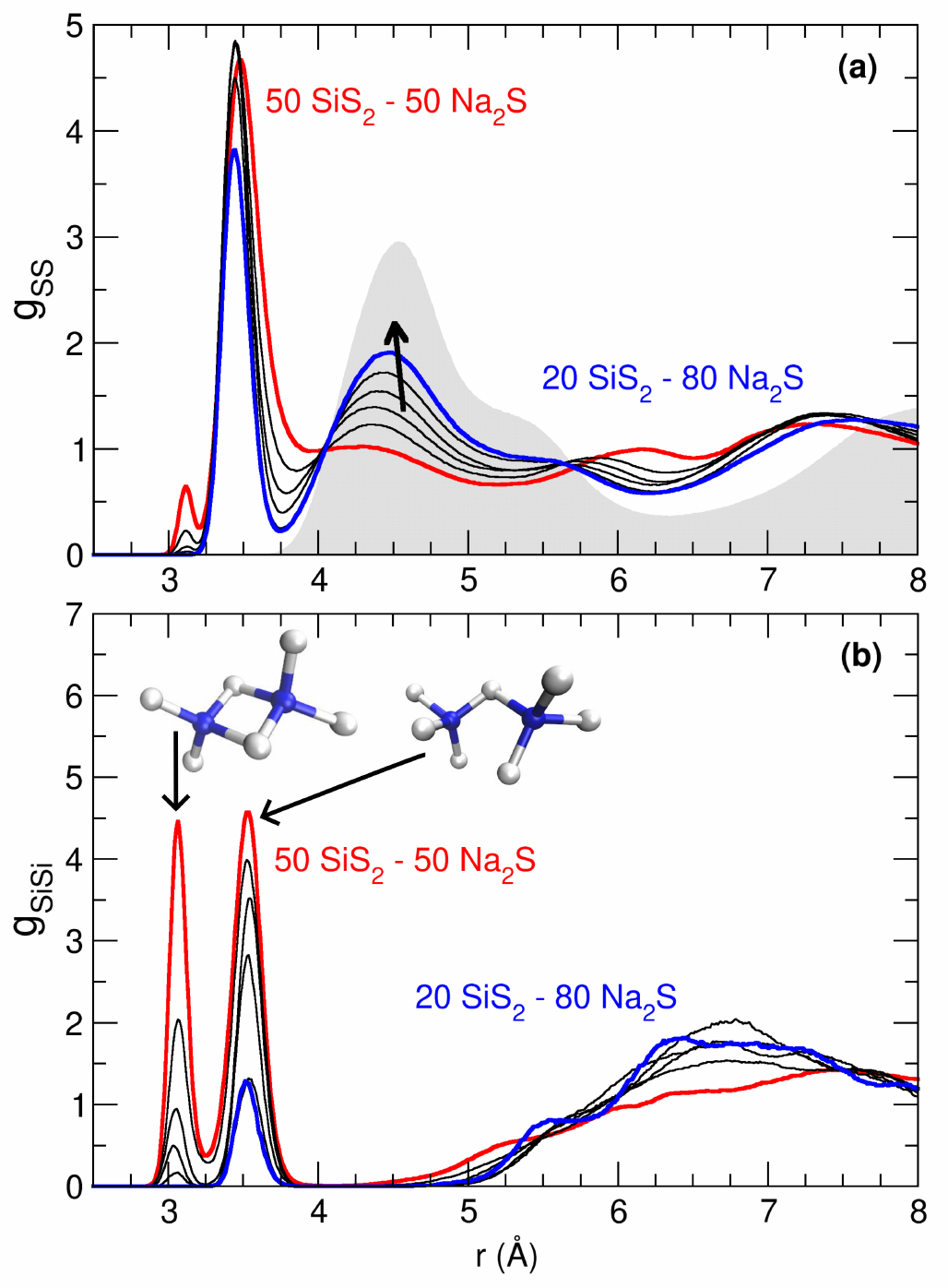}
\caption{\label{gij} Calculated pair correlation function $g_{SS}(r)$ (a) and $g_{SiSi}(r)$ (b) in glassy (300~K) $x$Na$_2$S -- (1-$x$)SiS$_2$ for different modifier content $x$ : 50 (red), 60, 66, 70, 75, 80 (blue). The gray zone in panel (a) corresponds to $g_{SS}(r)$ of pure Na$_2$S ($x$=100~\%). In panel b, the structural motifs (ES and CS) are associated with the principal peaks. Arrows indicate the peak evolution with increasing $x$.
}
\end{figure}

\begin{figure}
\includegraphics[width=0.7\linewidth]{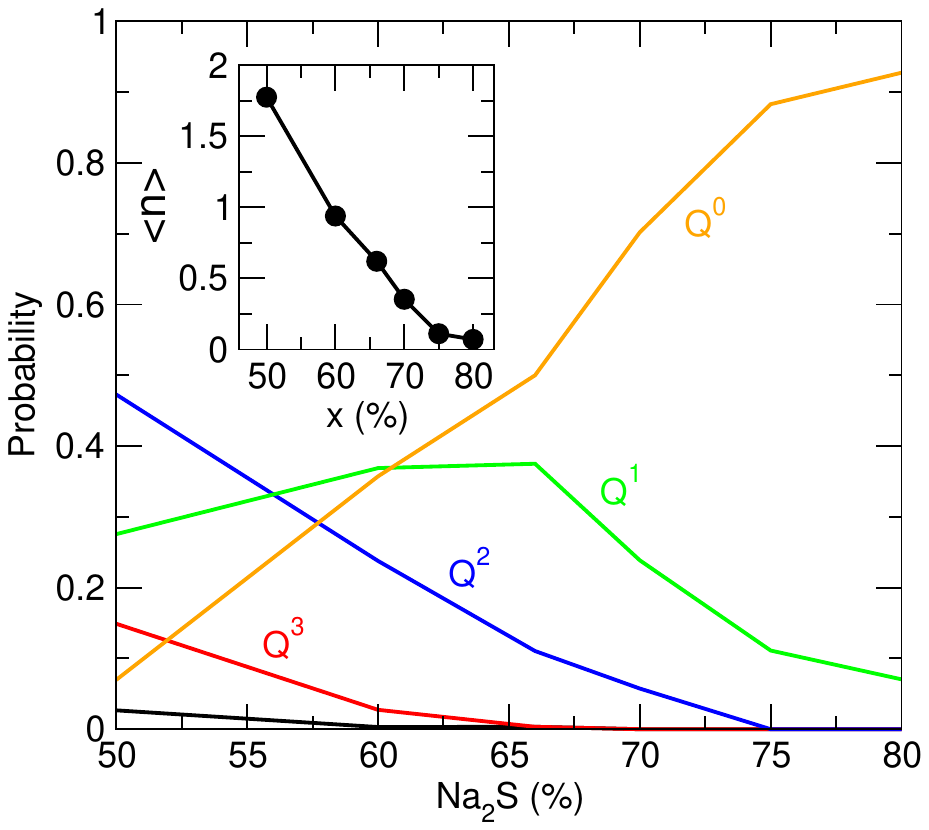}
\caption{\label{qn} Calculated population of $Q^n$ species as a function of modifier in $x$Na$_2$S -- (1-$x$)SiS$_2$ glasses. The black curve corresponds to $Q^4$ species. The inset represents the average number of BS atoms $\langle n\rangle$.
}
\end{figure}

In Fig. \ref{gij} we represent the S-S (a) and Si-Si (b) partial pair correlation functions for the different investigated compositions. Of special interest is the presence of prepeaks in both $g_{SS}(r)$ and $g_{SiSi}(r)$ in the region 3.0-3.2~\AA\ which signals the existence of ES structures as such motifs induce specific correlating distances\cite{salmon_jncs_2007,micoulaut_prb_2022,chakraborty_prb_2017}. The principal peak at 3.53~\AA\ in the Si-Si pair correlation function (panel b) corresponds to distances between Si centers of CS tetrahedra, whereas the one at 3.03~\AA\ is the Si-Si correlating distance involved in ES motifs. With increasing Na$_2$S content, the amplitude of such peaks reduces and signals a global breakdown of ES structures, consistently with experiments, that is induced by the reduction of tetrahedra connectivity and the increase of NBS atoms. A convenient way to quantify the depolymerization is to enumerate Si tetrahedra having $n$ BS atoms (or 4$-n$ NBS atoms) which is also known in the literature as the $Q^n$ speciation\cite{pradel_cras_2022}. Figure~\ref{qn} provides such an enumeration for the different considered glasses. Noteworthy is the fact that the 50Na$_2$S -- 50SiS$_2$ system contains a majority of $Q^2$ species which is consistent with the Na:Si stoichiometric ratio of 2:1, and this allows to have $Q^2$--$Q^2$ ES dimers. The ideal chemically ordered model\cite{mysen_book_2018} leads to 100~\% Q$^2$. With the growing proportion of $Q^1$ and $Q^0$ units upon Na$_2$S addition, ES tetrahedra become impossible as two BS atoms are needed to form ES connections so that their corresponding amount will decrease, consistently with the reduction of the principal peak at 3.03~\AA\ in $g_{SiSi}$ with growing modifier content $x$.

The trend of the different $Q^n$ species appears to be rather consistent with the overall observation of highly depolymerized network glasses such as the present system or the archetypal alkali silicates\cite{micoulaut_am_2008}, i.e. while the base SiS$_2$ network is made of 100~\% Q$^4$ species in ES or CS connection, at 50~\% modifier there is a distribution of possible tetrahedra, ranging from $Q^3$ (theoretically dominant at 33~\% modifier) to $Q^2$ (dominant at 50~\%), $Q^1$ (dominant at 60~\%) and $Q^0$ (dominant at 66~\%). The present calculated distribution and the trend with composition are globally compatible with a recent estimate from NMR, albeit our statistics does not fully reproduce the experimental results\cite{watson_jncs_2017} which consisted also in the presence of possible dimeric groupings containing a Si-Si homopolar bond (Na$_6$Si$_2$S$_6$). With increasing $x$, the global connectivity reduces as displayed in the inset (Fig. \ref{qn}) which represents the average tetrahedral connectivity $\langle n\rangle$ defined by :
\begin{eqnarray}
    \langle n\rangle =\sum\limits_{n}np_n
\end{eqnarray}
where $p_n$ is the calculated probability of finding a $Q^n$ species.

The other partial of interest (Fig. \ref{gij}a) represents sulfur-sulfur correlations with increasing Na$_2$S content. Here, we notice a prepeak at 3.1~\AA\ which is the signature of a small number of homopolar S--S bonds which are also detected in other modified sulfide glasses\cite{kassem_ic_2022}. The pair correlation function is dominated by a principal peak at 3.5~\AA\ which corresponds to the distance involved in the edge of the SiS$_{4/2}$ tetrahedra, and which is weakly sensitive to composition. Upon Na$_2$S addition, one acknowledges the emergence of a secondary peak at $r\simeq$4.5~\AA\ which corresponds to the growing presence of isolated Na$_2$S molecules. As a matter of fact, an investigation of pure glassy Na$_2$S (gray zone) indicates that the corresponding function $g_{SS}(r)$ has, indeed, a prominent peak at the same distance which corresponds to correlations between the sulfur centers of Na$_2$S molecules. For $x$=100~\%, the typical principal associated with the edges of the SiS$_{4/2}$ is, of course, absent.
\par

\begin{figure}
\includegraphics[width=0.7\linewidth]{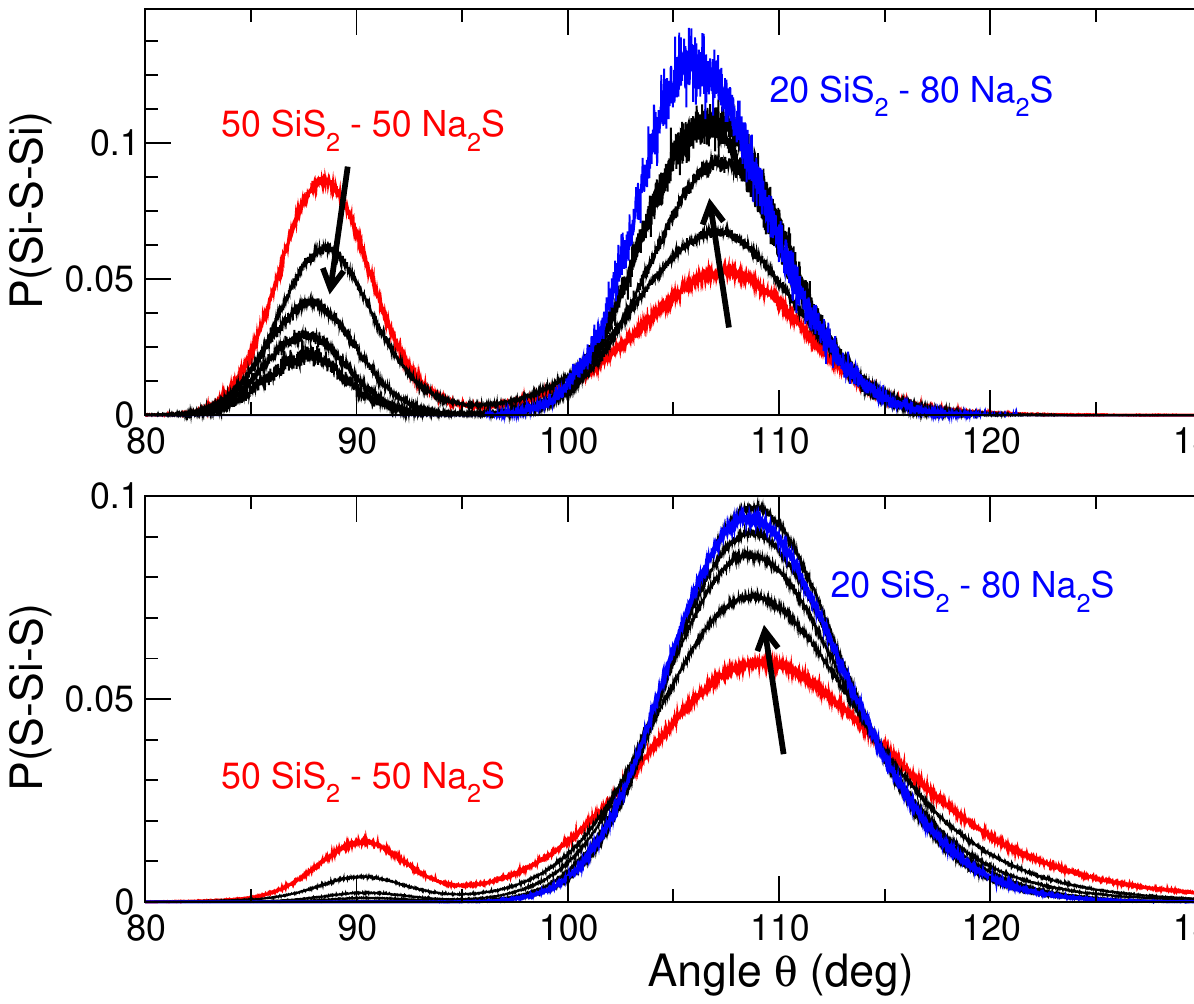}
\caption{\label{bad} Calculated bond angle distribution Si-S-Si (a) and S-Si-S (b) in glassy (300~K) $x$Na$_2$S -- (1-$x$)SiS$_2$ for different modifier content $x$ : 50 (red), 60, 66, 70, 75, 80 (blue).
}
\end{figure}

The vanishing of ES tetrahedra with growing $x$ can be also inferred from associated bond angle distributions (BAD, Fig. \ref{bad}) which display a bimodal structure typical of chalcogenide networks having both CS and ES connections\cite{chakraborty_prb_2017,micoulaut_prb_2022}. Here (Fig. \ref{bad}a), we acknowledge for the NS glass (red curve) two major contributions in the Si-S-Si BAD centred at 88$^\circ$ and 108$^\circ$ which can be identified with ES and CS tetrahedra, respectively. The same holds for the S-Si-S BAD (Fig. \ref{bad}b) which is dominated by the intra-tetrahedral angle at $\arccos(-1/3)$=109$^\circ$. With the addition of modifier, the ES contribution obviously vanish and, thus, also indicate a breakdown of such motifs.

\subsection {Emergence of the Na$_2$S phase} 

\begin{figure}
\includegraphics[width=0.7\linewidth]{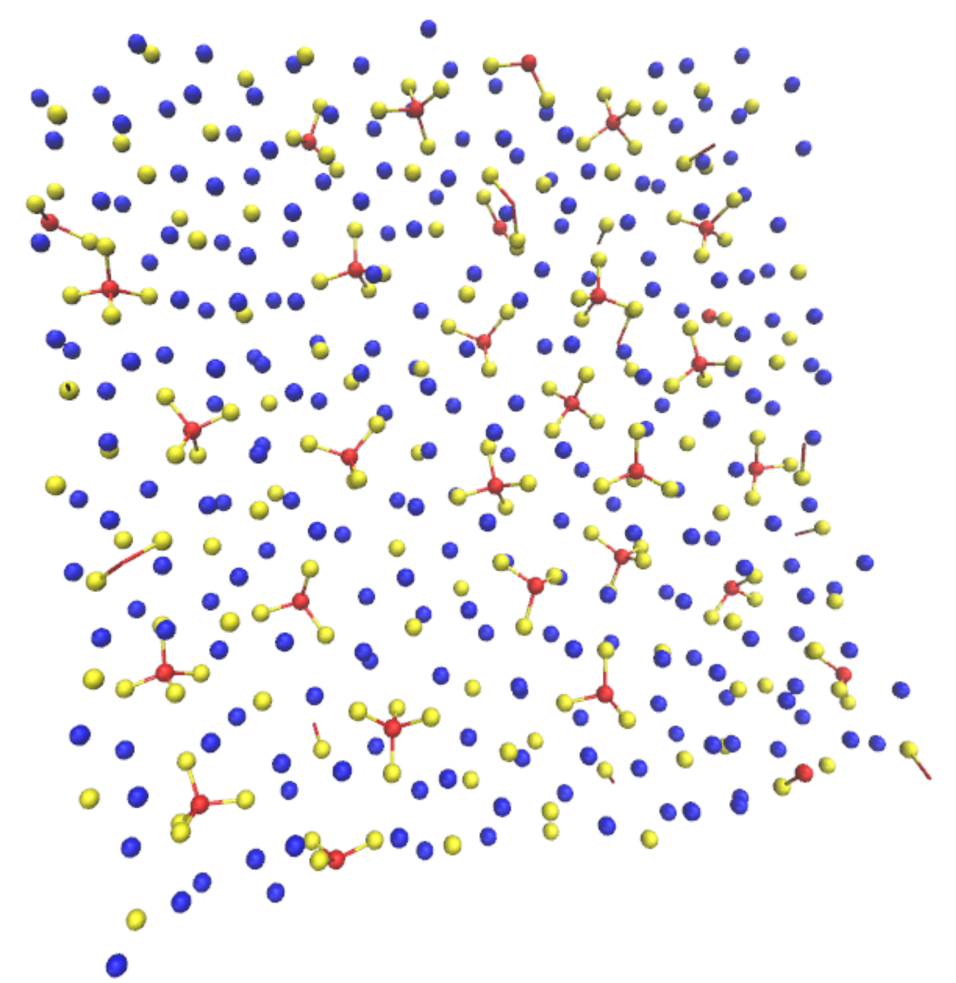}
\caption{\label{snap} Atomic snapshot of a slab (thickness 6~\AA) in a 300~K 80Na$_2$S -- 20SiS$_2$ glass made of Si (red), S (yellow) and Na (blue). Isolated sulfur atoms indicate the presence of Na$_2$S molecules.
}
\end{figure}

As mentioned earlier, with such extremely elevated concentrations of modifier, one is beyond the theoretical limit of having an entire $Q^0$ structure expected at 66~\% modifier, i.e. a network that is made of isolated Si tetrahedra. The precise account from the atomic scale trajectories is slightly different as we found a probability of 50.0, 37.5 and 11.0~\% for $Q^0$, Q$^1$ and $Q^2$ species, respectively (Fig. \ref{qn}), and similarly to various other MD studies\cite{du_jncs_2004,micoulaut_prb_2023} which also highlight the fact that there is a continuous conversion $Q^n\rightarrow Q^{n-1}$ with growing $x$. Upon additional modifier content however, a Na$_2$S liquid or glass phase is supposed to emerge as detected from atomic snapshots (Fig. \ref{snap}) which highlight the presence of isolated Si tetrahedra with a lot of sulfur and sodium atoms signaling the presence of Na$_2$S molecules. 

\begin{figure}
\includegraphics[width=0.7\linewidth]{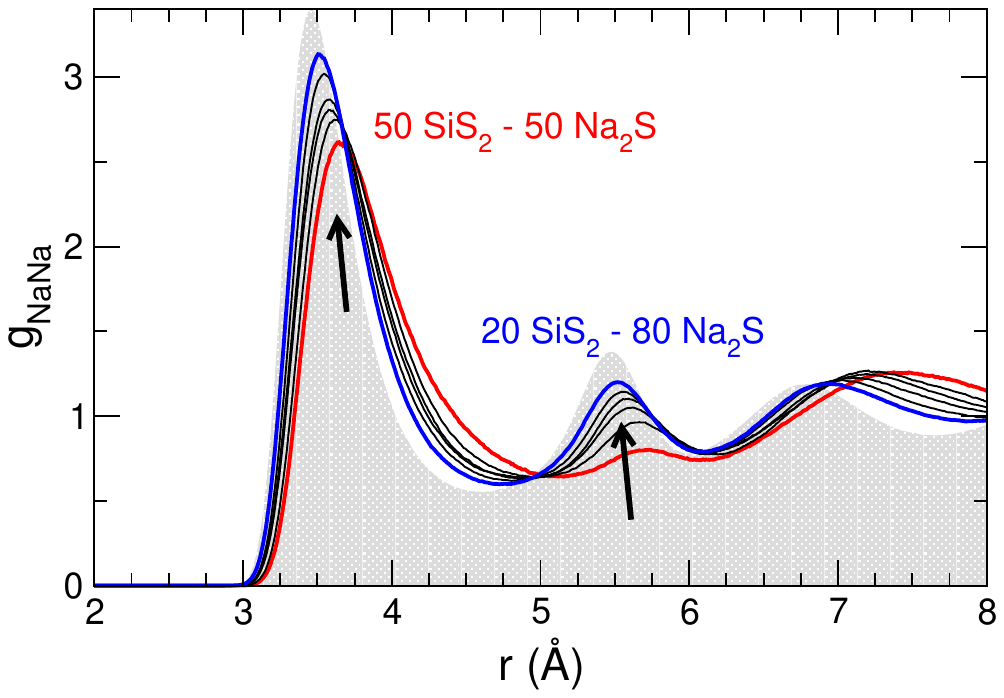}
\caption{\label{gijb} Calculated pair correlation function $g_{NaNa}(r)$ in glassy (300~K) $x$Na$_2$S -- (1-$x$)SiS$_2$ for different modifier content $x$ : 50 (red), 60, 66, 70, 75, 80 (blue). The gray zone corresponds to $g_{NaNa}(r)$ of pure Na$_2$S ($x$=100~\%). Arrows indicate the peak evolution with increasing $x$.
}
\end{figure}

The progressive emergence of correlations between Na$_2$S species is barely visible at the 50:50 composition in the Na-Na pair correlation function $g_{NaNa}(r)$ (Fig. \ref{gijb}) but secondary peaks emerge beyond the first coordination shell with increasing $x$, i.e. we detect at $r\simeq$5.5~\AA\ a distance whose amplitude grows with Na composition, and which appears to be typical of pure Na$_2$S (gray area).

\begin{figure}
\includegraphics[width=0.7\linewidth]{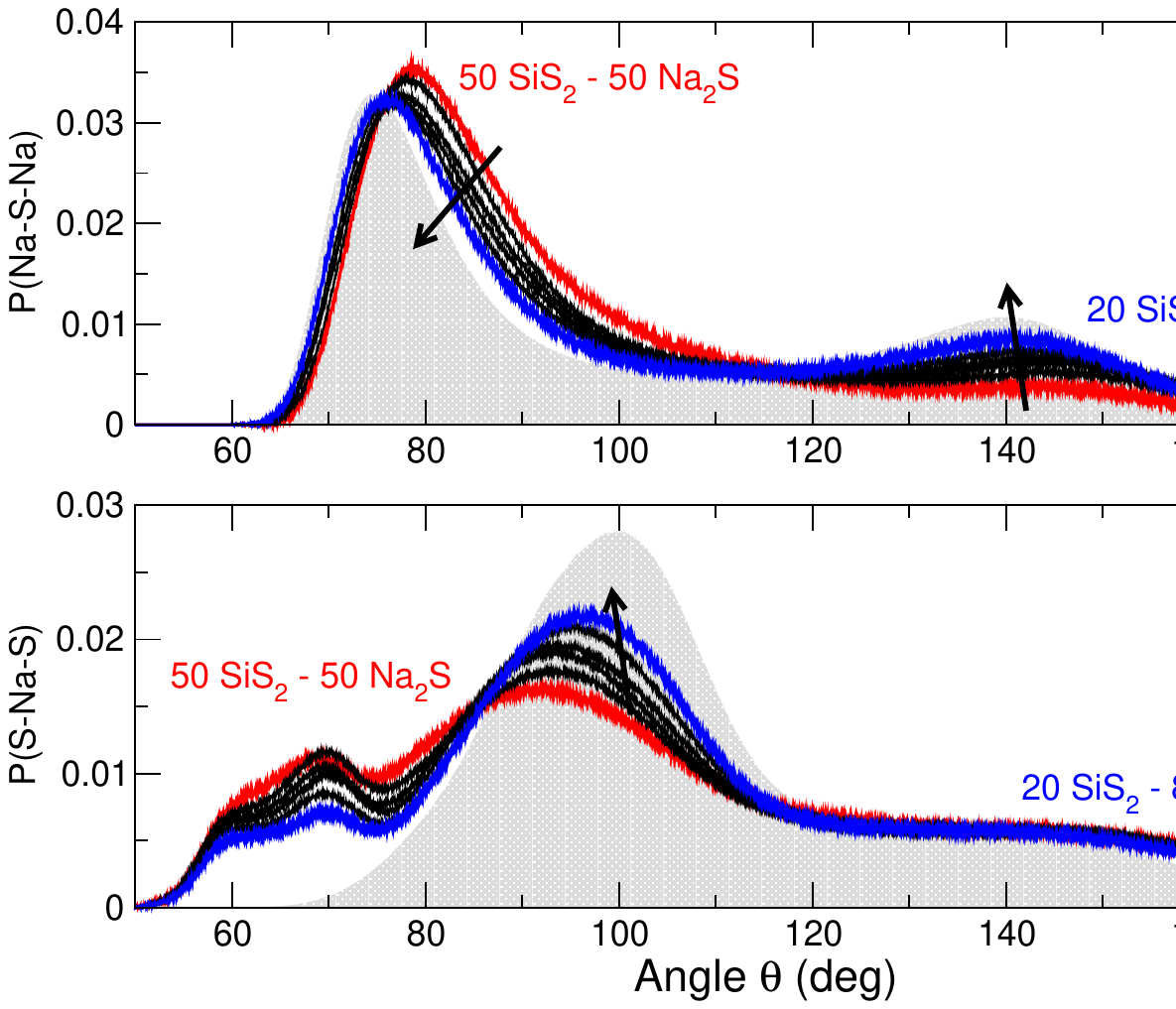}
\caption{\label{bad1} Calculated bond angle distribution Na-S-Na (a) and S-Na-S in glassy (300~K) $x$Na$_2$S -- (1-$x$)SiS$_2$ for different modifier content $x$ : 50 (red), 60, 66, 70, 75, 80 (blue). The gray zone corresponds to the bond angle distribution of pure Na$_2$S ($x$=100~\%).
}
\end{figure}

As crystalline Na$_2$S is in a cubic symmetry\cite{feher_zaac_1953,zintl_zeapc_1934}, for ultra-depolymerized glasses, we find short range features related to the crystalline order with typical angular contributions at 80-90$^\circ$ and 150-180$^\circ$ in the Na-S-Na and S-Na-S bond angle distributions (Fig. \ref{bad1}). For the former (panel a), such contributions grow significantly as we move from 50Na$_2$S -- 50SiS$_2$ to 80Na$_2$S -- 20SiS$_2$.

\subsection{Evidence for Na channel collapse}

The addition of Na atoms in the highly depolymerized systems, furthermore, leads to a vanishing of the channel dynamics that is typical of low modified oxides and sulfides. In the latter, the ion motion results in preferential pathways\cite{angell_jpc_1982,greaves_jncs_1985}, and these are evidenced both from simulations\cite{voigtmann_epl_2006,sunyer_prb_2002,bauchy_prb_2011,jund_prb_2001} and from quasi-elastic neutron scattering experiments\cite{meyer_prl_2004,meyer_nn_2012}. In the latter, a pre-peak is observed in the partial Na-Na structure factor $S_{NaNa}(k)$ at low momentum transfer $k_{PP}$=0.9~\AA$^{-1}$. This underscores the possibility of having sodium-rich regions with a characteristic length scale\cite{salmon_prsa_1994} of $\simeq$7~\AA\ embedded in a nearly frozen silica matrix with different compositions. These percolating channels appear with the Na clustering induced by the breakage of the inter-tetrahedral bond angle\cite{nesbitt_jncs_2015} and are already present at low modifier content\cite{jund_prb_2001}. While these effects are rather well documented in the archetypal alkali silicates, recently the possibility of such channels has been evidenced in sulfide glasses as well\cite{micoulaut_jncs_2024,sorensen_jpcb_2023}. 

\begin{figure}
\includegraphics[width=0.7\linewidth]{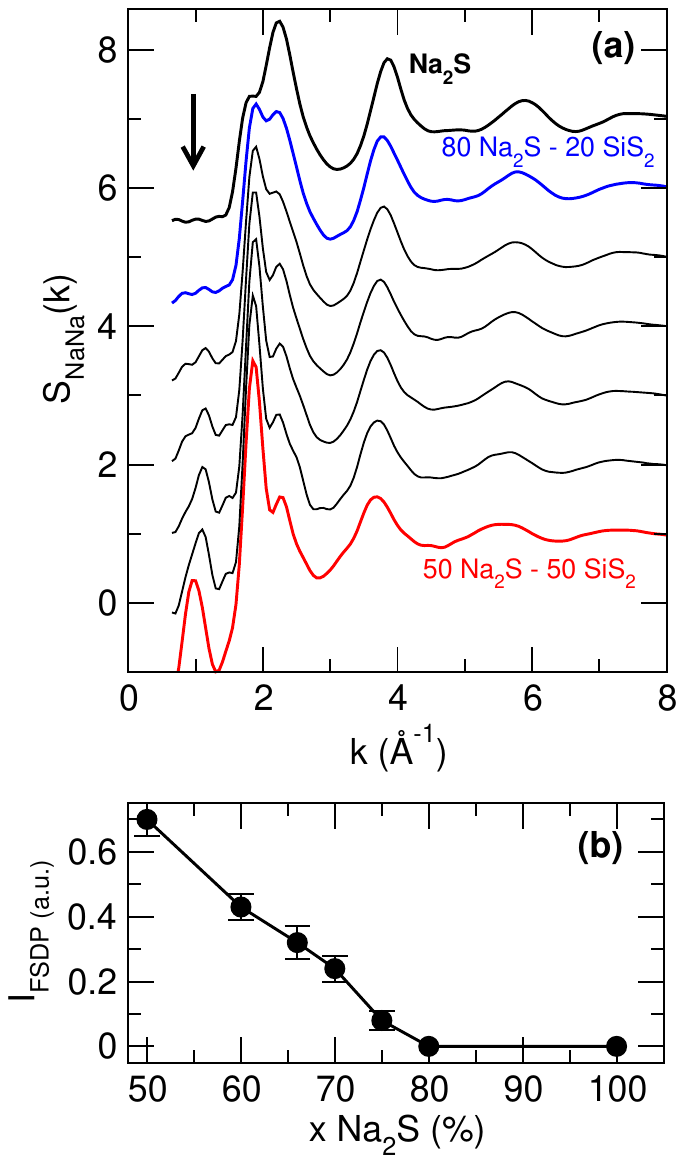}
\caption{\label{sqnana} (a) Calculated partial Na-Na structure factor $S_{NaNa}(k)$ at 300~K for the different investigated systems. The arrow indicates the FSDP region. (b) Intensity of the FSDP as a function of modifier content.
}
\end{figure}

Here, we find a similar feature with a clear and obvious signature of a pre-peak FSDP in the Na-Na partial structure factor (Fig. \ref{sqnana}a) that is observable at $k_{FSDP}\simeq$1.0~\AA$^{-1}$ for the NS system. The existence of this pre-peak is the signature\cite{meyer_prl_2004} that some ordering of Na ions takes place in the NS glass, and involves a typical length scale\cite{salmon_prsa_1994} of 7.7/$k_{FSDP}\simeq$7-8~\AA\ that is found to slightly shift to lower distance (higher $k$) with increasing modifier content. Interestingly, the intensity of the FSDP decreases dramatically as one moves from 50Na$_2$S -- 50SiS$_2$ to 20Na$_2$S -- 80SiS$_2$ and for $x\geq$75~\% the FSDP is barely visible. An estimate of its intensity $I_{FSDP}$, furthermore, indicates a linear decrease down to zero for $x$=80~\% (Fig. \ref{sqnana}b). 
\par
These features clearly suggest that in highly depolymerized systems, the Na channels vanish with the overall reduction of the network connectivity ensuring structural regions dominated by the (Si,S) connected species. As a result, one expects the nature of the dynamics to evolve from a dynamics within channels for $x\simeq$50~\% with a certain lengthscale to a spatially unconstrained diffusion at $x\simeq$80~\% typical of high temperature melts when the FSDP is absent. A recent inspection of atomic snapshots\cite{micoulaut_jncs_2024} has shown that the degree of network polymerization impacts the localization of the Na motion. At "low" $x$, the latter occurs within regions limited by the presence of a partially remaining (Si,S) network, whereas for a complete depolymerization, the Na motion almost extends to the entire simulation box so that the ion paths result from series of independent random segments as for Brownian motion which underscores the stochastic nature of the Na motion.


\subsection{Li versus Na}
\begin{figure}
\includegraphics[width=0.7\linewidth]{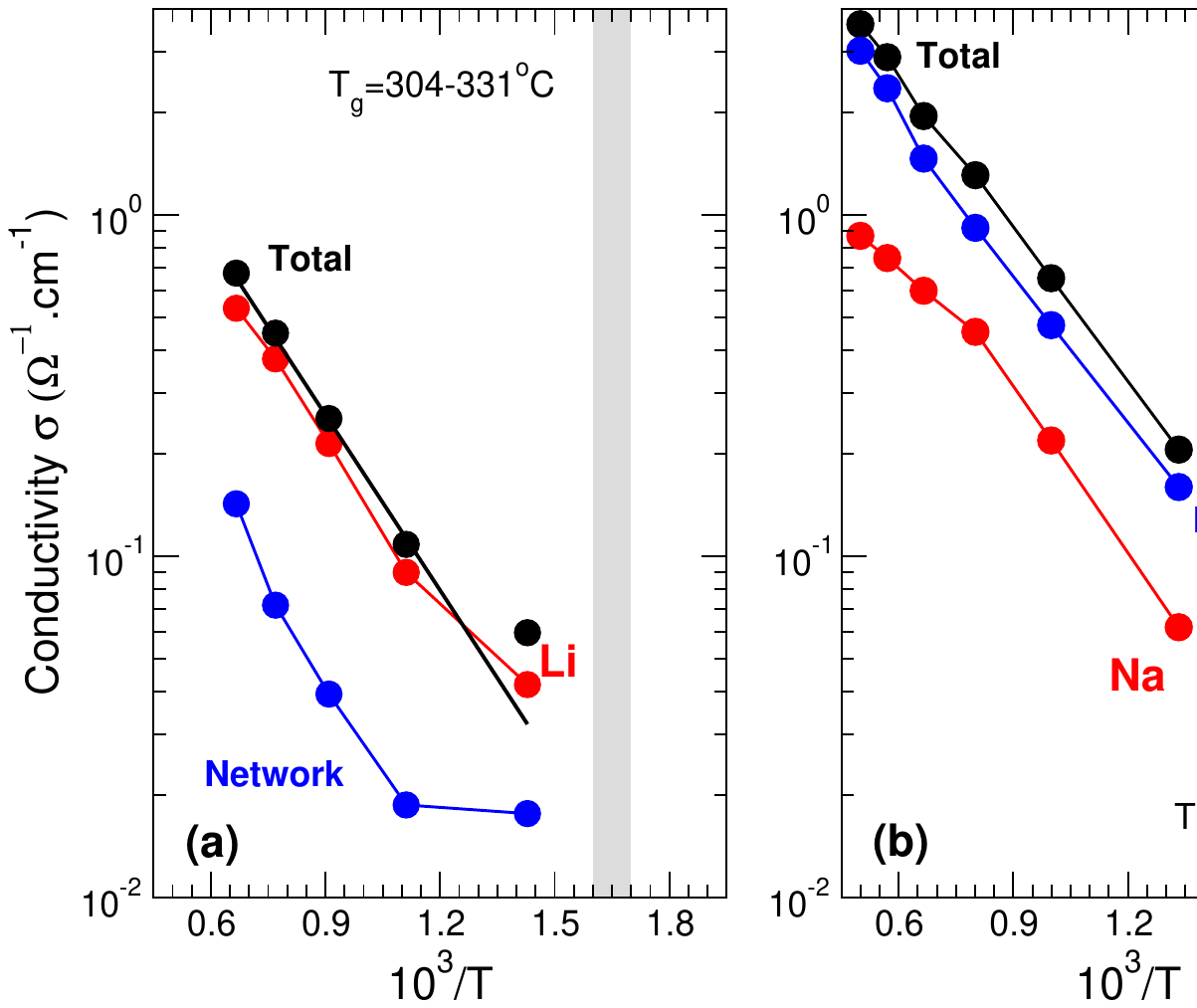}
\caption{\label{Na_vs_Li} Comparison between conductivity in liquid 60 M$_2$S -- 40SiS$_2$ with M=Li (a) or Na (b) and decomposition into contributions from the alkali ions (red) and network species (S,Si, blue). The gray zone corresponds to the respective glass transition region which is of about 304-331$^\circ$C in Li thiosilicates\cite{pradel_cras_2022} and 270$^\circ$C\cite{watson_jpcb_2018}.
}
\end{figure}

It is instructive to compare two isochemical compounds having the same amount of modifier content. In Fig. \ref{Na_vs_Li}b we represent the calculated conductivity $\sigma$ for the 60Na$_2$S-40SiS$_2$ that we compare with the corresponding lithium compound\cite{micoulaut_jacers_2024} (Fig. \ref{Na_vs_Li}a).

Both panels summarize what we have inferred above from the diffusivity difference $\Delta D$ between the alkali ion and the network-forming species (inset of Fig. \ref{diff2}). In the lithium compound, the conductivity is essentially controlled by the Li ions, the contribution of the network species being a factor 5 smaller for all considered temperatures, including for the lowest considered temperature (10$^3$/T=1.4, i.e. $T$=700~K). Conversely, given the rather limited diffusivity difference $\Delta D$ in the Na-glass, we find that $\sigma$ is essentially controlled by the (Si,S) atoms. If we discard still possible effects due the different (Li and Na) force-fields, our results indicates that any project of the electrolyte conductivity enhancement in these materials must target the network properties, and their underlying dynamics. An interesting perspective in this context might build on molecular flexibility which has been found to boost ionic conduction\cite{novita_prl_2007}. The reduction of the network connectivity in certain glasses drives, indeed, a rigid to flexible transition\cite{micoulaut_mrsb_2017} that brings into the glass flexible (or floppy) modes which reduce the energy barriers for local deformations that are then able to enhance the ion mobility\cite{micoulaut_prb_2009}. Recently, the link between rigidity and conductivity has been cast in a more rigorous approach\cite{micoulaut_prl_2025} which has shown that molecular flexibility promotes the occurrence of flexible modes and topological degrees of freedom in the network structure. These lead to a substantial increase of conductivity in the flexible phase of glasses\cite{boolchand_pssb_2018}. It suggests that molecular flexibility can serve as an efficient way for conductivity enhancement in all solid-state batteries using Na-based amorphous electrolytes.

\section{Summary and conclusion}
Here we have investigated the dynamic and electric behavior of highly depolymerized glassy and liquid networks having a large amount of modifier content. Prior to the investigation, we have established a force-field able to describe correctly the structure of glassy 50Na$_2$S -- 50SiS$_2$ (NS) in real and reciprocal space, and this effort represents a clear step forward with respect to previous attempts\cite{dive_jpcb_2018,sorensen_jpcb_2023}. The sodium thiosilicates have become increasingly attractive because of their possible use in fast ion batteries and because of the availability of sodium. In such ultra-depolymerized $x$Na$_2$S -- (1-$x$)SiS$_2$ networks, we find that for $x\geq$50~\% the increase of conductivity with modifier content is weak, and results from a too small diffusivity difference between the Na ion and the network species. This clearly contrasts with Li-based glasses whose electric properties are dominated by the very mobile Li ions. Given these results, the design of all solid state batteries using Na-based glassy electrolytes must therefore seriously consider the role of the network species. 
\par
The sodium thiosilicates represent a very interesting base material that can be used for further alloying with either other network formers (GeS$_2$ or P$_2$S$_5$, the mixed former effect) or other alkali modifiers (Li$_2$S, the mixed alkali effect) in the search for a continuous improvement of ion conduction. Our results clearly indicate a highly depolymerized network that is predominantly tetrahedral in character with the salient phenomeneology already encountered in archetypal modified (thio)silicate glasses, here a base network disruption (SiS$_2$) upon Na addition that leads to the growing presence of non-bridging sulfur having in its vicinity Na atoms, a $Q^n$ population that ultimately is dominated by $n$=0 at elevated modifier content (80~\% Na$_2$S), and by the growing Na$_2$S phase. 

\section*{Acknowledgments}
The authors acknowledges support from Chaire d'Excellence Sorbonne Université - Universidad Aut\'onoma de Mexico, and from Fondation MAIF pour la recherche. MM acknowledges continuous support from CNRS and Sorbonne Université, and has abandoned the idea of being funded by the controversial ANR. Interactions and discussions with Jean-Gabriel Barthélemy, Gerardo Naumis, Andrea Piarristeguy, Virginie Viallet, and Hugo Flores-Ruiz are also acknowledged.


\begin{mcitethebibliography}{58}
	\providecommand*\natexlab[1]{#1}
	\providecommand*\mciteSetBstSublistMode[1]{}
	\providecommand*\mciteSetBstMaxWidthForm[2]{}
	\providecommand*\mciteBstWouldAddEndPuncttrue
	{\def\EndOfBibitem{\unskip.}}
	\providecommand*\mciteBstWouldAddEndPunctfalse
	{\let\EndOfBibitem\relax}
	\providecommand*\mciteSetBstMidEndSepPunct[3]{}
	\providecommand*\mciteSetBstSublistLabelBeginEnd[3]{}
	\providecommand*\EndOfBibitem{}
	\mciteSetBstSublistMode{f}
	\mciteSetBstMaxWidthForm{subitem}{(\alph{mcitesubitemcount})}
	\mciteSetBstSublistLabelBeginEnd
	{\mcitemaxwidthsubitemform\space}
	{\relax}
	{\relax}
	
	\bibitem[Tarascon and Armand(2001)Tarascon, and Armand]{tarascon_nature_2001}
	Tarascon,~J.-M.; Armand,~M. Issues and challenges facing rechargeable lithium
	batteries. \emph{Nature} \textbf{2001}, \emph{414}, 359--367\relax
	\mciteBstWouldAddEndPuncttrue
	\mciteSetBstMidEndSepPunct{\mcitedefaultmidpunct}
	{\mcitedefaultendpunct}{\mcitedefaultseppunct}\relax
	\EndOfBibitem
	\bibitem[Weiss \latin{et~al.}(2021)Weiss, Ruess, Kasnatscheew, Levartovsky,
	Levy, Minnmann, Stolz, Waldmann, Wohlfahrt-Mehrens, Aurbach, \latin{et~al.}
	others]{weiss_aem_2021}
	Weiss,~M.; Ruess,~R.; Kasnatscheew,~J.; Levartovsky,~Y.; Levy,~N.~R.;
	Minnmann,~P.; Stolz,~L.; Waldmann,~T.; Wohlfahrt-Mehrens,~M.; Aurbach,~D.;
	others Fast charging of lithium-ion batteries: a review of materials aspects.
	\emph{Advanced Energy Materials} \textbf{2021}, \emph{11}, 2101126\relax
	\mciteBstWouldAddEndPuncttrue
	\mciteSetBstMidEndSepPunct{\mcitedefaultmidpunct}
	{\mcitedefaultendpunct}{\mcitedefaultseppunct}\relax
	\EndOfBibitem
	\bibitem[Huang \latin{et~al.}(2024)Huang, Chan, Wong, Liang, Sun, Wu, Lu, Lu,
	and Chen]{huang_ce_2024}
	Huang,~B.; Chan,~C.~H.; Wong,~H.~H.; Liang,~S.; Sun,~M.; Wu,~T.; Lu,~Q.;
	Lu,~L.; Chen,~B. Electrolyte Developments for All-Solid-State Lithium
	Batteries: Classifications, Recent Advances and Synthesis Methods.
	\emph{Batteries \& Supercaps} \textbf{2024}, \emph{n/a}, e202400432\relax
	\mciteBstWouldAddEndPuncttrue
	\mciteSetBstMidEndSepPunct{\mcitedefaultmidpunct}
	{\mcitedefaultendpunct}{\mcitedefaultseppunct}\relax
	\EndOfBibitem
	\bibitem[Grady \latin{et~al.}(2020)Grady, Wilkinson, Randall, and
	Mauro]{grady_frontiers_2020}
	Grady,~Z.~A.; Wilkinson,~C.~J.; Randall,~C.~A.; Mauro,~J.~C. Emerging role of
	non-crystalline electrolytes in solid-state battery research. \emph{Frontiers
		in Energy Research} \textbf{2020}, \emph{8}, 218\relax
	\mciteBstWouldAddEndPuncttrue
	\mciteSetBstMidEndSepPunct{\mcitedefaultmidpunct}
	{\mcitedefaultendpunct}{\mcitedefaultseppunct}\relax
	\EndOfBibitem
	\bibitem[Pradel and Piarristeguy(2022)Pradel, and
	Piarristeguy]{pradel_cras_2022}
	Pradel,~A.; Piarristeguy,~A. Thio and selenosilicates, sulfide and selenide
	counterparts of silicates: similarities and differences. \emph{Comptes
		Rendus. G{\'e}oscience} \textbf{2022}, \emph{354}, 79--99\relax
	\mciteBstWouldAddEndPuncttrue
	\mciteSetBstMidEndSepPunct{\mcitedefaultmidpunct}
	{\mcitedefaultendpunct}{\mcitedefaultseppunct}\relax
	\EndOfBibitem
	\bibitem[Eckert \latin{et~al.}(1990)Eckert, Zhang, and Kennedy]{eckert_cm_1990}
	Eckert,~H.; Zhang,~Z.; Kennedy,~J.~H. Structural transformation of non-oxide
	chalcogenide glasses. The short-range order of lithium sulfide
	(Li$_2$S)-phosphorus pentasulfide (P$_2$S$_5$) glasses studied by
	quantitative phosphorus-31, lithium-6, and lithium-7 high-resolution
	solid-state NMR. \emph{Chemistry of Materials} \textbf{1990}, \emph{2},
	273--279\relax
	\mciteBstWouldAddEndPuncttrue
	\mciteSetBstMidEndSepPunct{\mcitedefaultmidpunct}
	{\mcitedefaultendpunct}{\mcitedefaultseppunct}\relax
	\EndOfBibitem
	\bibitem[Mizuno \latin{et~al.}(2005)Mizuno, Hayashi, Tadanaga, and
	Tatsumisago]{mizuno_am_2005}
	Mizuno,~F.; Hayashi,~A.; Tadanaga,~K.; Tatsumisago,~M. New, highly
	ion-conductive crystals precipitated from Li$_2$S--P$_2$S$_5$ glasses.
	\emph{Advanced Materials} \textbf{2005}, \emph{17}, 918--921\relax
	\mciteBstWouldAddEndPuncttrue
	\mciteSetBstMidEndSepPunct{\mcitedefaultmidpunct}
	{\mcitedefaultendpunct}{\mcitedefaultseppunct}\relax
	\EndOfBibitem
	\bibitem[Mori \latin{et~al.}(2013)Mori, Ichida, Iwase, Otomo, Kohara, Arai,
	Uchimoto, Ogumi, Onodera, and Fukunaga]{mori_cpl_2013}
	Mori,~K.; Ichida,~T.; Iwase,~K.; Otomo,~T.; Kohara,~S.; Arai,~H.; Uchimoto,~Y.;
	Ogumi,~Z.; Onodera,~Y.; Fukunaga,~T. Visualization of conduction pathways in
	lithium superionic conductors: Li$_2$S-P$_2$S$_5$ glasses and
	Li$_7$P$_3$S$_{11}$ glass--ceramic. \emph{Chemical Physics Letters}
	\textbf{2013}, \emph{584}, 113--118\relax
	\mciteBstWouldAddEndPuncttrue
	\mciteSetBstMidEndSepPunct{\mcitedefaultmidpunct}
	{\mcitedefaultendpunct}{\mcitedefaultseppunct}\relax
	\EndOfBibitem
	\bibitem[Ohara \latin{et~al.}(2016)Ohara, Mitsui, Mori, Onodera, Shiotani,
	Koyama, Orikasa, Murakami, Shimoda, Mori, \latin{et~al.}
	others]{ohara_sr_2016}
	Ohara,~K.; Mitsui,~A.; Mori,~M.; Onodera,~Y.; Shiotani,~S.; Koyama,~Y.;
	Orikasa,~Y.; Murakami,~M.; Shimoda,~K.; Mori,~K.; others Structural and
	electronic features of binary Li$_2$S-P$_2$S$_5$ glasses. \emph{Scientific
		reports} \textbf{2016}, \emph{6}, 1--9\relax
	\mciteBstWouldAddEndPuncttrue
	\mciteSetBstMidEndSepPunct{\mcitedefaultmidpunct}
	{\mcitedefaultendpunct}{\mcitedefaultseppunct}\relax
	\EndOfBibitem
	\bibitem[Pradel \latin{et~al.}(1995)Pradel, Taillades, Ribes, and
	Eckert]{pradel_jncs_1995}
	Pradel,~A.; Taillades,~G.; Ribes,~M.; Eckert,~H. $^{29}$Si NMR structural
	studies of ionically conductive silicon chalcogenide glasses and model
	compounds. \emph{Journal of non-crystalline solids} \textbf{1995},
	\emph{188}, 75--86\relax
	\mciteBstWouldAddEndPuncttrue
	\mciteSetBstMidEndSepPunct{\mcitedefaultmidpunct}
	{\mcitedefaultendpunct}{\mcitedefaultseppunct}\relax
	\EndOfBibitem
	\bibitem[Micoulaut(2016)]{micoulaut_ropp_2016}
	Micoulaut,~M. Relaxation and physical aging in network glasses: a review.
	\emph{Reports on Progress in Physics} \textbf{2016}, \emph{79}, 066504\relax
	\mciteBstWouldAddEndPuncttrue
	\mciteSetBstMidEndSepPunct{\mcitedefaultmidpunct}
	{\mcitedefaultendpunct}{\mcitedefaultseppunct}\relax
	\EndOfBibitem
	\bibitem[Micoulaut(2024)]{micoulaut_jncs_2024}
	Micoulaut,~M. Molecular dynamics simulations of SiS2-Li2S-LiI fast ion glasses:
	Increase of conductivity is driven by network atoms. \emph{Journal of
		Non-Crystalline Solids} \textbf{2024}, \emph{636}, 123017\relax
	\mciteBstWouldAddEndPuncttrue
	\mciteSetBstMidEndSepPunct{\mcitedefaultmidpunct}
	{\mcitedefaultendpunct}{\mcitedefaultseppunct}\relax
	\EndOfBibitem
	\bibitem[Micoulaut \latin{et~al.}(2023)Micoulaut, Piarristeguy, Masson,
	Poitras, Escalier, Kachmar, and Pradel]{micoulaut_prb_2023}
	Micoulaut,~M.; Piarristeguy,~A.; Masson,~O.; Poitras,~L.-M.; Escalier,~R.;
	Kachmar,~A.; Pradel,~A. Quantitative assessment of network depolymerization
	in archetypal superionic glasses and its relationship with ion conduction: A
	case study on Na$_2$S- GeS$_2$. \emph{Physical Review B} \textbf{2023},
	\emph{108}, 144205\relax
	\mciteBstWouldAddEndPuncttrue
	\mciteSetBstMidEndSepPunct{\mcitedefaultmidpunct}
	{\mcitedefaultendpunct}{\mcitedefaultseppunct}\relax
	\EndOfBibitem
	\bibitem[Denoue \latin{et~al.}(2021)Denoue, David, Calers, Gautier, Verger, and
	Calvez]{denoue_mrb_2021}
	Denoue,~K.; David,~L.; Calers,~C.; Gautier,~A.; Verger,~L.; Calvez,~L. New
	synthesis route for glasses and glass-ceramics in the Ga$_2$S$_3$Na$_2$S
	binary system. \emph{Materials Research Bulletin} \textbf{2021}, \emph{142},
	111423\relax
	\mciteBstWouldAddEndPuncttrue
	\mciteSetBstMidEndSepPunct{\mcitedefaultmidpunct}
	{\mcitedefaultendpunct}{\mcitedefaultseppunct}\relax
	\EndOfBibitem
	\bibitem[Poitras and Micoulaut(2023)Poitras, and Micoulaut]{poitras_prb_2023}
	Poitras,~L.-M.; Micoulaut,~M. Establishment of an empirical force-field for
	crystalline and amorphous Li$_2$S- SiS$_2$ electrolytes. \emph{Physical
		Review B} \textbf{2023}, \emph{107}, 214205\relax
	\mciteBstWouldAddEndPuncttrue
	\mciteSetBstMidEndSepPunct{\mcitedefaultmidpunct}
	{\mcitedefaultendpunct}{\mcitedefaultseppunct}\relax
	\EndOfBibitem
	\bibitem[Watson and Martin(2018)Watson, and Martin]{watson_jpcb_2018}
	Watson,~D.~E.; Martin,~S.~W. Composition Dependence of the Glass-Transition
	Temperature and Molar Volume in Sodium Thiosilicophosphate Glasses: A
	Structural Interpretation Using a Real Solution Model. \emph{The Journal of
		Physical Chemistry B} \textbf{2018}, \emph{122}, 10637--10646\relax
	\mciteBstWouldAddEndPuncttrue
	\mciteSetBstMidEndSepPunct{\mcitedefaultmidpunct}
	{\mcitedefaultendpunct}{\mcitedefaultseppunct}\relax
	\EndOfBibitem
	\bibitem[Wright(1993)]{wright_jncs_1993}
	Wright,~A.~C. The comparison of molecular dynamics simulations with diffraction
	experiments. \emph{Journal of Non-Crystalline Solids} \textbf{1993},
	\emph{159}, 264--268\relax
	\mciteBstWouldAddEndPuncttrue
	\mciteSetBstMidEndSepPunct{\mcitedefaultmidpunct}
	{\mcitedefaultendpunct}{\mcitedefaultseppunct}\relax
	\EndOfBibitem
	\bibitem[Dive \latin{et~al.}(2018)Dive, Benmore, Wilding, Martin, Beckman, and
	Banerjee]{dive_jpcb_2018}
	Dive,~A.; Benmore,~C.; Wilding,~M.; Martin,~S.; Beckman,~S.; Banerjee,~S.
	Molecular dynamics modeling of the structure and Na+-ion transport in
	Na$_2$S+SiS$_2$ glassy electrolytes. \emph{The Journal of Physical Chemistry
		B} \textbf{2018}, \emph{122}, 7597--7608\relax
	\mciteBstWouldAddEndPuncttrue
	\mciteSetBstMidEndSepPunct{\mcitedefaultmidpunct}
	{\mcitedefaultendpunct}{\mcitedefaultseppunct}\relax
	\EndOfBibitem
	\bibitem[Flores-Ruiz \latin{et~al.}(2018)Flores-Ruiz, Micoulaut, Piarristeguy,
	Coulet, Johnson, Cuello, and Pradel]{flores_prb_2018}
	Flores-Ruiz,~H.; Micoulaut,~M.; Piarristeguy,~A.; Coulet,~M.-V.; Johnson,~M.;
	Cuello,~G.; Pradel,~A. Structural, vibrational, and dynamic properties of
	Ge-Ga-Te liquids with increasing connectivity: A combined neutron scattering
	and molecular dynamics study. \emph{Physical Review B} \textbf{2018},
	\emph{97}, 214207\relax
	\mciteBstWouldAddEndPuncttrue
	\mciteSetBstMidEndSepPunct{\mcitedefaultmidpunct}
	{\mcitedefaultendpunct}{\mcitedefaultseppunct}\relax
	\EndOfBibitem
	\bibitem[Bertani \latin{et~al.}(2022)Bertani, Pallini, Cocchi, Menziani, and
	Pedone]{bertani_jacers_2022}
	Bertani,~M.; Pallini,~A.; Cocchi,~M.; Menziani,~M.~C.; Pedone,~A. A new
	self-consistent empirical potential model for multicomponent borate and
	borosilicate glasses. \emph{Journal of the American Ceramic Society}
	\textbf{2022}, \emph{105}, 7254--7271\relax
	\mciteBstWouldAddEndPuncttrue
	\mciteSetBstMidEndSepPunct{\mcitedefaultmidpunct}
	{\mcitedefaultendpunct}{\mcitedefaultseppunct}\relax
	\EndOfBibitem
	\bibitem[Bertani \latin{et~al.}(2021)Bertani, Menziani, and
	Pedone]{bertani_prm_2021}
	Bertani,~M.; Menziani,~M.~C.; Pedone,~A. Improved empirical force field for
	multicomponent oxide glasses and crystals. \emph{Phys. Rev. Mater.}
	\textbf{2021}, \emph{5}, 045602\relax
	\mciteBstWouldAddEndPuncttrue
	\mciteSetBstMidEndSepPunct{\mcitedefaultmidpunct}
	{\mcitedefaultendpunct}{\mcitedefaultseppunct}\relax
	\EndOfBibitem
	\bibitem[Chakraborty \latin{et~al.}(2017)Chakraborty, Boolchand, and
	Micoulaut]{chakraborty_prb_2017}
	Chakraborty,~S.; Boolchand,~P.; Micoulaut,~M. Structural properties of Ge-S
	amorphous networks in relationship with rigidity transitions: An ab initio
	molecular dynamics study. \emph{Physical Review B} \textbf{2017}, \emph{96},
	094205\relax
	\mciteBstWouldAddEndPuncttrue
	\mciteSetBstMidEndSepPunct{\mcitedefaultmidpunct}
	{\mcitedefaultendpunct}{\mcitedefaultseppunct}\relax
	\EndOfBibitem
	\bibitem[Micoulaut \latin{et~al.}(2022)Micoulaut, Pethes, J{\'o}v{\'a}ri,
	Pusztai, Krbal, W{\'a}gner, Prokop, Michalik, Ikeda, and
	Kaban]{micoulaut_prb_2022}
	Micoulaut,~M.; Pethes,~I.; J{\'o}v{\'a}ri,~P.; Pusztai,~L.; Krbal,~M.;
	W{\'a}gner,~T.; Prokop,~V.; Michalik,~{\v{S}}.; Ikeda,~K.; Kaban,~I.
	Structural properties of chalcogenide glasses and the isocoordination rule:
	Disentangling effects from chemistry and network topology. \emph{Physical
		Review B} \textbf{2022}, \emph{106}, 014206\relax
	\mciteBstWouldAddEndPuncttrue
	\mciteSetBstMidEndSepPunct{\mcitedefaultmidpunct}
	{\mcitedefaultendpunct}{\mcitedefaultseppunct}\relax
	\EndOfBibitem
	\bibitem[S{\o}rensen \latin{et~al.}(2023)S{\o}rensen, Smedskjaer, and
	Micoulaut]{sorensen_jpcb_2023}
	S{\o}rensen,~S.~S.; Smedskjaer,~M.~M.; Micoulaut,~M. Evidence for Complex
	Dynamics in Glassy Fast Ion Conductors: The Case of Sodium Thiosilicates.
	\emph{The Journal of Physical Chemistry B} \textbf{2023}, \emph{127},
	10179--10188\relax
	\mciteBstWouldAddEndPuncttrue
	\mciteSetBstMidEndSepPunct{\mcitedefaultmidpunct}
	{\mcitedefaultendpunct}{\mcitedefaultseppunct}\relax
	\EndOfBibitem
	\bibitem[Thomas \latin{et~al.}(1985)Thomas, Peterson, and
	Hutchinson]{thomas_jacer_1985}
	Thomas,~M.; Peterson,~N.~L.; Hutchinson,~E. Tracer Diffusion and Electrical
	Conductivity in Sodium-Rubidium Silicon Sulfide Glasses. \emph{Journal of the
		American Ceramic Society} \textbf{1985}, \emph{68}, 99--104\relax
	\mciteBstWouldAddEndPuncttrue
	\mciteSetBstMidEndSepPunct{\mcitedefaultmidpunct}
	{\mcitedefaultendpunct}{\mcitedefaultseppunct}\relax
	\EndOfBibitem
	\bibitem[Hansen and McDonald(2013)Hansen, and McDonald]{hansen_book}
	Hansen,~J.-P.; McDonald,~I.~R. \emph{Theory of simple liquids: with
		applications to soft matter}; Academic press, 2013\relax
	\mciteBstWouldAddEndPuncttrue
	\mciteSetBstMidEndSepPunct{\mcitedefaultmidpunct}
	{\mcitedefaultendpunct}{\mcitedefaultseppunct}\relax
	\EndOfBibitem
	\bibitem[Ribes \latin{et~al.}(1980)Ribes, Barrau, and Souquet]{ribes_jncs_1980}
	Ribes,~M.; Barrau,~B.; Souquet,~J. Sulfide glasses: Glass forming region,
	structure and ionic conduction of glasses in Na$_2$S-XS$_2$ (X= Si; Ge),
	Na$_2$S- P$_2$S$_5$ and Li$_2$S- GeS$_2$ systems. \emph{Journal of
		Non-Crystalline Solids} \textbf{1980}, \emph{38}, 271--276\relax
	\mciteBstWouldAddEndPuncttrue
	\mciteSetBstMidEndSepPunct{\mcitedefaultmidpunct}
	{\mcitedefaultendpunct}{\mcitedefaultseppunct}\relax
	\EndOfBibitem
	\bibitem[Golodnitsky \latin{et~al.}(2015)Golodnitsky, Strauss, Peled, and
	Greenbaum]{golodnitsky_jes_2015}
	Golodnitsky,~D.; Strauss,~E.; Peled,~E.; Greenbaum,~S. On order and disorder in
	polymer electrolytes. \emph{Journal of The Electrochemical Society}
	\textbf{2015}, \emph{162}, A2551\relax
	\mciteBstWouldAddEndPuncttrue
	\mciteSetBstMidEndSepPunct{\mcitedefaultmidpunct}
	{\mcitedefaultendpunct}{\mcitedefaultseppunct}\relax
	\EndOfBibitem
	\bibitem[Tu \latin{et~al.}(2022)Tu, Veith, Butt, and Floudas]{tu_m_2022}
	Tu,~C.-H.; Veith,~L.; Butt,~H.-J.; Floudas,~G. Ionic Conductivity of a Solid
	Polymer Electrolyte Confined in Nanopores. \emph{Macromolecules}
	\textbf{2022}, \emph{55}, 1332--1341\relax
	\mciteBstWouldAddEndPuncttrue
	\mciteSetBstMidEndSepPunct{\mcitedefaultmidpunct}
	{\mcitedefaultendpunct}{\mcitedefaultseppunct}\relax
	\EndOfBibitem
	\bibitem[Aniya and Ikeda(2020)Aniya, and Ikeda]{aniya_pssb_2020}
	Aniya,~M.; Ikeda,~M. Arrhenius Crossover Phenomena and Ionic Conductivity in
	Ionic Glass-Forming Liquids. \emph{physica status solidi (b)} \textbf{2020},
	\emph{257}, 2000139\relax
	\mciteBstWouldAddEndPuncttrue
	\mciteSetBstMidEndSepPunct{\mcitedefaultmidpunct}
	{\mcitedefaultendpunct}{\mcitedefaultseppunct}\relax
	\EndOfBibitem
	\bibitem[Malki \latin{et~al.}(2015)Malki, Magnien, Pinet, and
	Richet]{malki_gca_2015}
	Malki,~M.; Magnien,~V.; Pinet,~O.; Richet,~P. Electrical conductivity of
	iron-bearing silicate glasses and melts. Implications for the mechanisms of
	iron redox reactions. \emph{Geochimica et Cosmochimica Acta} \textbf{2015},
	\emph{165}, 137--147\relax
	\mciteBstWouldAddEndPuncttrue
	\mciteSetBstMidEndSepPunct{\mcitedefaultmidpunct}
	{\mcitedefaultendpunct}{\mcitedefaultseppunct}\relax
	\EndOfBibitem
	\bibitem[Fan \latin{et~al.}(2020)Fan, Del~Campo, Montouillout, and
	Malki]{fan_jncs_2020}
	Fan,~H.; Del~Campo,~L.; Montouillout,~V.; Malki,~M. Ionic conductivity and
	boron anomaly in binary lithium borate melts. \emph{Journal of
		Non-Crystalline Solids} \textbf{2020}, \emph{543}, 120160\relax
	\mciteBstWouldAddEndPuncttrue
	\mciteSetBstMidEndSepPunct{\mcitedefaultmidpunct}
	{\mcitedefaultendpunct}{\mcitedefaultseppunct}\relax
	\EndOfBibitem
	\bibitem[Watson and Martin(2017)Watson, and Martin]{watson_jncs_2017}
	Watson,~D.~E.; Martin,~S.~W. Short range order characterization of the Na$_2$S+
	SiS$_2$ glass system using Raman, infrared and 29Si magic angle spinning
	nuclear magnetic resonance spectroscopies. \emph{Journal of Non-Crystalline
		Solids} \textbf{2017}, \emph{471}, 39--50\relax
	\mciteBstWouldAddEndPuncttrue
	\mciteSetBstMidEndSepPunct{\mcitedefaultmidpunct}
	{\mcitedefaultendpunct}{\mcitedefaultseppunct}\relax
	\EndOfBibitem
	\bibitem[Dietrich \latin{et~al.}(2017)Dietrich, Weber, Sedlmaier, Indris,
	Culver, Walter, Janek, and Zeier]{dietrich_jmca_2017}
	Dietrich,~C.; Weber,~D.~A.; Sedlmaier,~S.~J.; Indris,~S.; Culver,~S.~P.;
	Walter,~D.; Janek,~J.; Zeier,~W.~G. Lithium ion conductivity in
	Li$_2$S--P$_2$S$_5$ glasses--building units and local structure evolution
	during the crystallization of superionic conductors Li$_3$PS$_4$,
	Li$_7$P$_3$S$_{11}$ and Li$_4$P$_2$S$_7$. \emph{Journal of Materials
		Chemistry A} \textbf{2017}, \emph{5}, 18111--18119\relax
	\mciteBstWouldAddEndPuncttrue
	\mciteSetBstMidEndSepPunct{\mcitedefaultmidpunct}
	{\mcitedefaultendpunct}{\mcitedefaultseppunct}\relax
	\EndOfBibitem
	\bibitem[Micoulaut \latin{et~al.}(2024)Micoulaut, Poitras, S{\o}rensen,
	Flores-Ruiz, and Naumis]{micoulaut_jacers_2024}
	Micoulaut,~M.; Poitras,~L.-M.; S{\o}rensen,~S.~S.; Flores-Ruiz,~H.;
	Naumis,~G.~G. Compressibility, diffusivity, and elasticity in relationship
	with ionic conduction: An atomic scale description of densified
	Li$_2$S--SiS$_2$ glasses. \emph{Journal of the American Ceramic Society}
	\textbf{2024}, \emph{107}, 7711--7726\relax
	\mciteBstWouldAddEndPuncttrue
	\mciteSetBstMidEndSepPunct{\mcitedefaultmidpunct}
	{\mcitedefaultendpunct}{\mcitedefaultseppunct}\relax
	\EndOfBibitem
	\bibitem[Eckert \latin{et~al.}(1989)Eckert, Kennedy, Pradel, and
	Ribes]{eckert_jncs_1989}
	Eckert,~H.; Kennedy,~J.~H.; Pradel,~A.; Ribes,~M. Structural transformation of
	thiosilicate glasses: $^{29}$Si MAS-NMR evidence for edge-sharing in the
	system Li$_2$S - SiS$_2$. \emph{Journal of non-crystalline solids}
	\textbf{1989}, \emph{113}, 287--293\relax
	\mciteBstWouldAddEndPuncttrue
	\mciteSetBstMidEndSepPunct{\mcitedefaultmidpunct}
	{\mcitedefaultendpunct}{\mcitedefaultseppunct}\relax
	\EndOfBibitem
	\bibitem[Salmon(2007)]{salmon_jncs_2007}
	Salmon,~P.~S. Structure of liquids and glasses in the Ge--Se binary system.
	\emph{Journal of Non-Crystalline Solids} \textbf{2007}, \emph{353},
	2959--2974\relax
	\mciteBstWouldAddEndPuncttrue
	\mciteSetBstMidEndSepPunct{\mcitedefaultmidpunct}
	{\mcitedefaultendpunct}{\mcitedefaultseppunct}\relax
	\EndOfBibitem
	\bibitem[Mysen and Richet(2018)Mysen, and Richet]{mysen_book_2018}
	Mysen,~B.; Richet,~P. \emph{Silicate glasses and melts}; Elsevier, 2018\relax
	\mciteBstWouldAddEndPuncttrue
	\mciteSetBstMidEndSepPunct{\mcitedefaultmidpunct}
	{\mcitedefaultendpunct}{\mcitedefaultseppunct}\relax
	\EndOfBibitem
	\bibitem[Micoulaut(2008)]{micoulaut_am_2008}
	Micoulaut,~M. Amorphous materials: Properties, structure, and durability:
	Constrained interactions, rigidity, adaptative networks, and their role for
	the description of silicates. \emph{American Mineralogist} \textbf{2008},
	\emph{93}, 1732--1748\relax
	\mciteBstWouldAddEndPuncttrue
	\mciteSetBstMidEndSepPunct{\mcitedefaultmidpunct}
	{\mcitedefaultendpunct}{\mcitedefaultseppunct}\relax
	\EndOfBibitem
	\bibitem[Kassem \latin{et~al.}(2022)Kassem, Bounazef, Sokolov, Bokova,
	Fontanari, Hannon, Alekseev, and Bychkov]{kassem_ic_2022}
	Kassem,~M.; Bounazef,~T.; Sokolov,~A.; Bokova,~M.; Fontanari,~D.;
	Hannon,~A.~C.; Alekseev,~I.; Bychkov,~E. Deciphering Fast Ion Transport in
	Glasses: A Case Study of Sodium and Silver Vitreous Sulfides. \emph{Inorganic
		Chemistry} \textbf{2022}, \emph{61}, 12870--12885\relax
	\mciteBstWouldAddEndPuncttrue
	\mciteSetBstMidEndSepPunct{\mcitedefaultmidpunct}
	{\mcitedefaultendpunct}{\mcitedefaultseppunct}\relax
	\EndOfBibitem
	\bibitem[Du and Cormack(2004)Du, and Cormack]{du_jncs_2004}
	Du,~J.; Cormack,~A. The medium range structure of sodium silicate glasses: a
	molecular dynamics simulation. \emph{Journal of Non-Crystalline Solids}
	\textbf{2004}, \emph{349}, 66--79\relax
	\mciteBstWouldAddEndPuncttrue
	\mciteSetBstMidEndSepPunct{\mcitedefaultmidpunct}
	{\mcitedefaultendpunct}{\mcitedefaultseppunct}\relax
	\EndOfBibitem
	\bibitem[Feh{\'e}r and Berthold(1953)Feh{\'e}r, and Berthold]{feher_zaac_1953}
	Feh{\'e}r,~F.; Berthold,~H. Beitr{\"a}ge zur Chemie des Schwefels. XIV.
	{\"U}ber das System Natrium—Schwefel. \emph{Zeitschrift f{\"u}r
		anorganische und allgemeine Chemie} \textbf{1953}, \emph{273}, 144--160\relax
	\mciteBstWouldAddEndPuncttrue
	\mciteSetBstMidEndSepPunct{\mcitedefaultmidpunct}
	{\mcitedefaultendpunct}{\mcitedefaultseppunct}\relax
	\EndOfBibitem
	\bibitem[Zintl \latin{et~al.}(1934)Zintl, Harder, and Dauth]{zintl_zeapc_1934}
	Zintl,~E.; Harder,~A.; Dauth,~B. Gitterstruktur der oxyde, sulfide, selenide
	und telluride des lithiums, natriums und kaliums. \emph{Zeitschrift f{\"u}r
		Elektrochemie und angewandte physikalische Chemie} \textbf{1934}, \emph{40},
	588--593\relax
	\mciteBstWouldAddEndPuncttrue
	\mciteSetBstMidEndSepPunct{\mcitedefaultmidpunct}
	{\mcitedefaultendpunct}{\mcitedefaultseppunct}\relax
	\EndOfBibitem
	\bibitem[Angell \latin{et~al.}(1982)Angell, Cheeseman, and
	Tamaddon]{angell_jpc_1982}
	Angell,~C.; Cheeseman,~P.; Tamaddon,~S. Computer simulation studies of
	migration mechanisms in ionic glasses and liquids. \emph{Le Journal de
		Physique Colloques} \textbf{1982}, \emph{43}, C9--381\relax
	\mciteBstWouldAddEndPuncttrue
	\mciteSetBstMidEndSepPunct{\mcitedefaultmidpunct}
	{\mcitedefaultendpunct}{\mcitedefaultseppunct}\relax
	\EndOfBibitem
	\bibitem[Greaves(1985)]{greaves_jncs_1985}
	Greaves,~G. EXAFS and the structure of glass. \emph{Journal of Non-Crystalline
		Solids} \textbf{1985}, \emph{71}, 203--217\relax
	\mciteBstWouldAddEndPuncttrue
	\mciteSetBstMidEndSepPunct{\mcitedefaultmidpunct}
	{\mcitedefaultendpunct}{\mcitedefaultseppunct}\relax
	\EndOfBibitem
	\bibitem[Voigtmann and Horbach(2006)Voigtmann, and Horbach]{voigtmann_epl_2006}
	Voigtmann,~T.; Horbach,~J. Slow dynamics in ion-conducting sodium silicate
	melts: Simulation and mode-coupling theory. \emph{Europhysics Letters}
	\textbf{2006}, \emph{74}, 459\relax
	\mciteBstWouldAddEndPuncttrue
	\mciteSetBstMidEndSepPunct{\mcitedefaultmidpunct}
	{\mcitedefaultendpunct}{\mcitedefaultseppunct}\relax
	\EndOfBibitem
	\bibitem[Sunyer \latin{et~al.}(2002)Sunyer, Jund, and Jullien]{sunyer_prb_2002}
	Sunyer,~E.; Jund,~P.; Jullien,~R. Characterization of channel diffusion in a
	sodium tetrasilicate glass via molecular-dynamics simulations. \emph{Physical
		Review B} \textbf{2002}, \emph{65}, 214203\relax
	\mciteBstWouldAddEndPuncttrue
	\mciteSetBstMidEndSepPunct{\mcitedefaultmidpunct}
	{\mcitedefaultendpunct}{\mcitedefaultseppunct}\relax
	\EndOfBibitem
	\bibitem[Bauchy and Micoulaut(2011)Bauchy, and Micoulaut]{bauchy_prb_2011}
	Bauchy,~M.; Micoulaut,~M. From pockets to channels: Density-controlled
	diffusion in sodium silicates. \emph{Physical Review B} \textbf{2011},
	\emph{83}, 184118\relax
	\mciteBstWouldAddEndPuncttrue
	\mciteSetBstMidEndSepPunct{\mcitedefaultmidpunct}
	{\mcitedefaultendpunct}{\mcitedefaultseppunct}\relax
	\EndOfBibitem
	\bibitem[Jund \latin{et~al.}(2001)Jund, Kob, and Jullien]{jund_prb_2001}
	Jund,~P.; Kob,~W.; Jullien,~R. Channel diffusion of sodium in a silicate glass.
	\emph{Physical Review B} \textbf{2001}, \emph{64}, 134303\relax
	\mciteBstWouldAddEndPuncttrue
	\mciteSetBstMidEndSepPunct{\mcitedefaultmidpunct}
	{\mcitedefaultendpunct}{\mcitedefaultseppunct}\relax
	\EndOfBibitem
	\bibitem[Meyer \latin{et~al.}(2004)Meyer, Horbach, Kob, Kargl, and
	Schober]{meyer_prl_2004}
	Meyer,~A.; Horbach,~J.; Kob,~W.; Kargl,~F.; Schober,~H. Channel formation and
	intermediate range order in sodium silicate melts and glasses. \emph{Physical
		Review Letters} \textbf{2004}, \emph{93}, 027801\relax
	\mciteBstWouldAddEndPuncttrue
	\mciteSetBstMidEndSepPunct{\mcitedefaultmidpunct}
	{\mcitedefaultendpunct}{\mcitedefaultseppunct}\relax
	\EndOfBibitem
	\bibitem[Meyer \latin{et~al.}(2012)Meyer, Kargl, and Horbach]{meyer_nn_2012}
	Meyer,~A.; Kargl,~F.; Horbach,~J. Channel diffusion in sodium silicate melts.
	\emph{Neutron News} \textbf{2012}, \emph{23}, 35--37\relax
	\mciteBstWouldAddEndPuncttrue
	\mciteSetBstMidEndSepPunct{\mcitedefaultmidpunct}
	{\mcitedefaultendpunct}{\mcitedefaultseppunct}\relax
	\EndOfBibitem
	\bibitem[Salmon(1994)]{salmon_prsa_1994}
	Salmon,~P.~S. Real space manifestation of the first sharp diffraction peak in
	the structure factor of liquid and glassy materials. \emph{Proceedings of the
		Royal Society of London. Series A: Mathematical and Physical Sciences}
	\textbf{1994}, \emph{445}, 351--365\relax
	\mciteBstWouldAddEndPuncttrue
	\mciteSetBstMidEndSepPunct{\mcitedefaultmidpunct}
	{\mcitedefaultendpunct}{\mcitedefaultseppunct}\relax
	\EndOfBibitem
	\bibitem[Nesbitt \latin{et~al.}(2015)Nesbitt, Henderson, Bancroft, and
	Ho]{nesbitt_jncs_2015}
	Nesbitt,~H.; Henderson,~G.; Bancroft,~G.; Ho,~R. Experimental evidence for Na
	coordination to bridging oxygen in Na-silicate glasses: Implications for
	spectroscopic studies and for the modified random network model.
	\emph{Journal of Non-Crystalline Solids} \textbf{2015}, \emph{409},
	139--148\relax
	\mciteBstWouldAddEndPuncttrue
	\mciteSetBstMidEndSepPunct{\mcitedefaultmidpunct}
	{\mcitedefaultendpunct}{\mcitedefaultseppunct}\relax
	\EndOfBibitem
	\bibitem[Novita \latin{et~al.}(2007)Novita, Boolchand, Malki, and
	Micoulaut]{novita_prl_2007}
	Novita,~D.~I.; Boolchand,~P.; Malki,~M.; Micoulaut,~M. Fast-ion conduction and
	flexibility of glassy networks. \emph{Physical review letters} \textbf{2007},
	\emph{98}, 195501\relax
	\mciteBstWouldAddEndPuncttrue
	\mciteSetBstMidEndSepPunct{\mcitedefaultmidpunct}
	{\mcitedefaultendpunct}{\mcitedefaultseppunct}\relax
	\EndOfBibitem
	\bibitem[Micoulaut and Yue(2017)Micoulaut, and Yue]{micoulaut_mrsb_2017}
	Micoulaut,~M.; Yue,~Y. Material functionalities from molecular rigidity:
	Maxwell’s modern legacy. \emph{Mrs Bulletin} \textbf{2017}, \emph{42},
	18--22\relax
	\mciteBstWouldAddEndPuncttrue
	\mciteSetBstMidEndSepPunct{\mcitedefaultmidpunct}
	{\mcitedefaultendpunct}{\mcitedefaultseppunct}\relax
	\EndOfBibitem
	\bibitem[Micoulaut \latin{et~al.}(2009)Micoulaut, Malki, Novita, and
	Boolchand]{micoulaut_prb_2009}
	Micoulaut,~M.; Malki,~M.; Novita,~D.; Boolchand,~P. Fast-ion conduction and
	flexibility and rigidity of solid electrolyte glasses. \emph{Physical Review
		B} \textbf{2009}, \emph{80}, 184205\relax
	\mciteBstWouldAddEndPuncttrue
	\mciteSetBstMidEndSepPunct{\mcitedefaultmidpunct}
	{\mcitedefaultendpunct}{\mcitedefaultseppunct}\relax
	\EndOfBibitem
	\bibitem[Micoulaut(2025)]{micoulaut_prl_2025}
	Micoulaut,~M. Ionic conductivity in disordered media: Molecular flexibility as
	a new paradigm to enhance ion motion in glassy electrolytes. \emph{Physical
		review letters} \textbf{2025}, \emph{134}\relax
	\mciteBstWouldAddEndPuncttrue
	\mciteSetBstMidEndSepPunct{\mcitedefaultmidpunct}
	{\mcitedefaultendpunct}{\mcitedefaultseppunct}\relax
	\EndOfBibitem
	\bibitem[Boolchand \latin{et~al.}(2018)Boolchand, Bauchy, Micoulaut, and
	Yildirim]{boolchand_pssb_2018}
	Boolchand,~P.; Bauchy,~M.; Micoulaut,~M.; Yildirim,~C. Topological phases of
	chalcogenide glasses encoded in the melt dynamics. \emph{physica status
		solidi (b)} \textbf{2018}, \emph{255}, 1800027\relax
	\mciteBstWouldAddEndPuncttrue
	\mciteSetBstMidEndSepPunct{\mcitedefaultmidpunct}
	{\mcitedefaultendpunct}{\mcitedefaultseppunct}\relax
	\EndOfBibitem
\end{mcitethebibliography}
\providecommand{\latin}[1]{#1}
\makeatletter
\providecommand{\doi}
{\begingroup\let\do\@makeother\dospecials
	\catcode`\{=1 \catcode`\}=2 \doi@aux}
\providecommand{\doi@aux}[1]{\endgroup\texttt{#1}}
\makeatother
\providecommand*\mcitethebibliography{\thebibliography}
\csname @ifundefined\endcsname{endmcitethebibliography}
{\let\endmcitethebibliography\endthebibliography}{}

\bibliographystyle{achemso.bst}

\newpage
\begin{table}
    \centering
    \begin{ruledtabular}
    \begin{tabular}{ccccc}
    Atom $i$ &  Atom $j$ & $A_{ij}$ (eV)& $\rho_{ij}$ (\AA) & $C_{ij}$ (eV$^{-1}$.\AA$^{-6}$)  \\
    \colrule
    \\
    S & S &42800& 0.284 &30 \\
    S & Si & 164000&0.178&80\\
    S & Na &833000&0.182&70\\
    Si & Si &-&-&- \\
    Si & Na &-&-&- \\   
    Na & Na &13000&0.3&300\\
    
    \end{tabular}
    \caption{Buckingham parameters of eq. (\ref{buck}) that describe xNa$_2$S -- (1-x)SiS$_2$ glasses (x $\in$ $\{$0.5, 0.6, 0.66, 0.7, 0.75, 0.8, 1.0$\}$). Charges have been taken as $q_{Si}$=$z_{Si}e$=2.4$e$, $q_S$=$z_S$=-1.2$e$ and $q_{Na}$=0.6$e$. The interactions between Si-Si and Si-Na are assumed to be purely Coulombic for the sake of simplicity. For the three-body interactions, we used $k_b$=1.5 and $\theta_0$=60$^\circ$.}
    \label{tab_nasis}
    \end{ruledtabular}
\end{table}

\newpage

\newpage
\newpage

\newpage

\newpage

\newpage

\newpage
\newpage

\newpage
\newpage
\newpage

\newpage
\newpage

\newpage

\end{document}